\documentclass[runningheads]{llncs}

\usepackage{amssymb}
\usepackage{graphicx}
\usepackage{tikz}

\newcommand{\Mcal}{\mathcal{M}}
\newcommand{\Lcal}{\mathcal{L}}

\begin{document}

\begin{frontmatter}

\title{\Large Analogue-digital systems with modes of physical behaviour}

\author{Edwin Beggs\inst{1} \and John V.\ Tucker\inst{2}}

\institute{Department of Mathematics, College of Science, \\ Swansea University, Singleton Park, Swansea, SA2 8PP, \\ Wales, United Kingdom \\ \email{e.j.beggs@swansea.ac.uk}  \and Department of Computer Science, College of Science, \\ Swansea University, Singleton Park, Swansea, SA2 8PP, \\ Wales, United Kingdom \\ \email{j.v.tucker@swansea.ac.uk}}

\titlerunning{Analogue-digital systems with modes of physical behaviour}
\authorrunning{Beggs and Tucker}

\maketitle

\pagenumbering{arabic}

\begin{abstract}
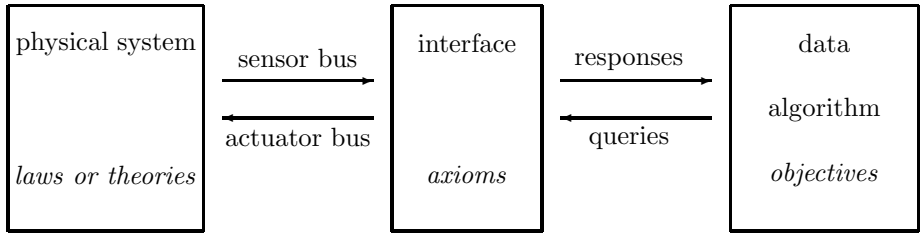
Complex environments, processes and systems may exhibit several distinct modes of physical behaviour or operation. Thus, for example, in their design, a \textit{set} of mathematical models may be needed, each model having its own domain of application and representing a particular mode of behaviour or operation of physical reality.  The models may be of disparate kinds -- discrete or continuous in data, time and space. Furthermore, some physical modes may not have a reliable model. Physical measurements  determine modes of operation. We explore the question: \textit{What is a mode of behaviour? How do we specify algorithms and software that monitor or govern a complex physical situation with many modes? How do we specify a portfolio of modes, and the computational problem of transitioning from using one mode to another mode as physical modes change?}  We propose a general definition of an \textit{analogue-digital system with modes}.  We show how any diverse set of modes -- with or without models -- can be bound together, and how the transitions between modes can be determined, by constructing a topological data type based upon a simplicial complex. We illustrate the ideas of physical modes and our theory by reflecting on simple examples, including driverless racing cars.
\end{abstract}

\end{frontmatter}

{ \quote
\textit{No plan survives contact with the enemy.} 
\begin{flushright}
After Helmuth von Moltke the Elder (1800-1891)
\end{flushright}
\endquote}

{ \quote
\textit{There are known knowns; there are things we know that we know.
There are known unknowns; that is to say, there are things that we now know we don't know.
But there are also unknown unknowns -- there are things we do not know we don't know.} 
\begin{flushright} Donald Rumsfeld (2002)
\end{flushright}
\endquote}

\section{Introduction} \label{vcgaksuy}

A typical analogue-digital system is a system in which a continuous physical environment, process or component is monitored or governed by a discrete algorithmic process. This simple description covers an astonishingly large range of systems, for example: classical control systems for machines, industrial plant, and buildings; networked products and vehicles; human monitoring and surveillance systems; and scientific experiments, mediated by software, that measure physical quantities. Many applications involve hybrid systems and cyber-physical systems\footnote{A cyber-physical system generalises the concept of an embedded system to a network of interacting devices with physical input and output \cite{Lee}. For example, cyber-physical systems and the internet of things are central to speculations on the future of manufacturing in \cite{INDUSTRIE}.} These are types of analogue-digital systems whose deployment is vast in scope and whose formal theories and design methods have matured; in particular, hybrid systems have a solid theoretical basis \cite{LunzeLamnabhi-Lagarrigue}, which we will discuss shortly (in subsection \ref{HybridSystemsIntro}). We will take a fresh look at such systems and raise and offer answers to some foundational questions for a general theory of analogue-digital systems, reflecting on the interface between the physical and the algorithmic, and defining formally a new class of complex analogue-digital systems having distinct modes of physical behaviour.

\subsection{Observations on real-world analogue-digital systems}

To begin, consider from first principles, a real-world analogue-digital system that consists of physical equipment that is controlled by software on a processor.  It is analogue-digital because the system involves continuous and discrete data. Typically, the data characterising the equipment is represented by real numbers, and the data processed by the program is represented by bits. The equipment is made from physical components that exchange data with the controlling software.  An important design criterion is the 
\newline
\newline
\textbf{Principle 0: Robustness.} \textit{The design of the software and its underlying algorithms must be able to cope with {\em any} data that may be output from, or input to, the physical components.}
\newline
\newline
Typically, the design of the software is based upon 

(a) mathematical models -- generally called \textit{dynamical systems} --  of the behaviour of the physical equipment; 

(b) exception handlers if none of the mathematical models apply to a mode of behaviour; 

(c) logical and algebraic models of the behaviour of the programs. 

\noindent Of course, in the case of (c), the behaviour of the programs is derived from the semantics of the specification and programming languages employed; we will not consider this factor. Consider (a): how the software depends on models of the behaviour of the equipment by, or derived from, dynamical systems. 

To create the software for an analogue-digital system, the physical must be replaced by an abstract specification that documents certain operations, tests and properties and that constitutes a data type interface between physical quantities and the algorithms and software.  The interface must enable portability and verification so that the software can be certified ``as reliable as possible''  according to some best practice engineering standard. For the designer of an analogue-digital system, specification and validation presents certain problems outside software engineering.  The reliability of an analogue-digital system depends upon the abstract assumptions about the physical system, the sensors and actuators. These assumptions are based on dynamical systems that model physical reality.

A dynamical system is a mathematical model of an entity, process, or environment whose behaviour changes in time. The model represents behaviour by means of states that change over time. The model is likely to have parameters that are external inputs, which represent either 

(i) external influences that are not predictable physical factors to which the system must respond or, conversely, 

(ii) known parameters or instructions designed to control the system.

\noindent The state of a dynamical system can be based on continuous data, discrete data or a combination of both. 

Complex environments, processes and systems may exhibit quite \textit{distinct} modes of physical behaviour or operation. Thus, for the design of complex physical environments, processes and systems, a \textit{set} of mathematical models may be needed, each model having its own physical domain of application, and representing a particular mode of behaviour or operation of physical reality. The models may be of disparate kinds. Certainly, no single mathematical model is adequate.  Furthermore, the models may not cover adequately all the possible modes of behaviour, i.e., not all physical modes may have a reliable model.  

\subsection{Modelling analogue-digital systems with modes}

This paper is about complex analogue-digital systems with many modes. We address the following questions: In a complex analogue-digital system,

\textit{What is a mode of behaviour?}

\textit{How can modes of physical behaviour, with a portfolio of different models, be specified for the use of algorithms and software?}

\textit{How can modes of physical behaviour, without sound models, be specified for the use of algorithms and software?}

\textit{How are all the modes bound or linked together to make a robust system? As the system operates, one model becomes less relevant and another more relevant model must be chosen to replace it. What conditions govern the transition from one mode to another?}

We introduce initial concept of an analogue-digital system of the form

\textit{physical system + interface and protocol for data exchange + algorithm}.

\noindent Specifically, we propose the algorithms and software treat the physical system as an \textit{oracle}, whose queries are mediated by a protocol governing an interface.\footnote{The idea is a generalisation of Turing's and Post's ideas about oracles to algorithms in computability theory.}  A good deal is known about the basic properties of such a physical oracle model in cases where the physical systems are \textit{very} simple and can be faithfully captured by a \textit{single} dynamical system and, therefore, have one and only one mode \cite{Oracles2,Oracles1,Wheatstone,Axiomatising,ImapctModels,ThreeExperiments}. Here we generalise this concept of an analogue-digital system by introducing modes that may have models or need to be handled as exceptional situations. See the schematic in Figure 1 with four modes.

\unitlength 0.6 mm


\smallskip
\smallskip
\smallskip
\smallskip

We propose a theoretical framework for the development of such analogue-digital systems with modes of physical behaviour and their evaluates their fitness.  Essentially, we define the concept of a mode to be a data type representing a physical behaviour and equipped with criteria that evaluate its accuracy or relevance.

To the interface and data exchange protocol are added a set of \textit{mode transition functions} that specify what needs to be done to make a transition from one mode to another.  We create a simple geometric description of the system, one which describes both the relationship between the various modes of operation of the system and when transitions between the modes are necessary. Such a geometry needs a topological space and so we construct a simplicial  complex -- a space made of points, lines, triangles, tetrahedra etc., glued together using simple rules -- that we call a \textit{nerve}. The points represent the modes, and the lines or triangles joining them are used to show how well each mode describes the physical system. As this description is very explicit, it forms a data type that can be used represent visually the mode transition protocol in the design of the algorithmic structure of the portfolio.  Working from first principles, we offer a new perspective that is distinct from hybrid systems.

\subsection{Hybrid systems}\label{HybridSystemsIntro}

Over the years, the term hybrid system has had a number of meanings, such as combining analogue and digital computation  \cite{TruHyb}. Currently, the term usually refers to a \textit{control} system with analogue and digital behaviour \cite{LunzeLamnabhi-Lagarrigue}, which is an interpretation suitable for our investigation here. Yet, as the ways of studying such systems are many and growing there is a reluctance to define the term (cf. \cite{LabinazGuay}). We prefer to speak informally of analogue-digital systems for which we are also studying general computability theories \cite{AnalogueDigitalModel,ChurchTuring,TZ2007,TZ2014}.  The computational aspects of designing hybrid systems received a great impetus in the 1990s from studies by Zohar Manna, Amir Pnueli and Tom Henzinger \cite{Maler,HenzHybAut}, who developed formal methods for modelling, based on finite automata and differential equations, and for reasoning, based on the use of modal and temporal logics and model checking. This has been a foundation for extensive research, tool building and applications in computer science. Thus, for some the term hybrid systems suggests work on control systems using these formal methods. We will summarise some features of the formal methods for modelling to help clarify the ways our approach is distinct; we return to these points in the concluding remarks  (\ref{HybridSystemsConclusion}). 

Hybrid systems are studied by petri nets or hybrid automata. In their graphical representation, the boxes in the hybrid automata are called \textit{locations} or \textit{finite states}.\footnote{Terminology varies in the literature -- sometimes the finite states in the automaton are called modes.}
 The boxes can have \textit{invariants} - conditions which must be satisfied as long as the system is in that state, and if they stop being satisfied, then the system must move to another state. The arrows between the boxes represent the allowed state transitions and have \textit{guard conditions}, which must be satisfied for the system to move along that arrow. The arrows can also have associated \textit{actions}, which specify instantaneous (or nearly so) changes of variables on taking the arrow. The states typically have equations for continuous changes of variables, e.g.\ differential equations. In some implementations we can have hierarchies of finite states, where opening the box for one finite state reveals another hybrid automaton inside: Ptolemy \cite{LeeTriPtol} takes such a hierarchical approach with high and low levels of description (the top level components are referred to as modes).
 
To contrast this with what we are proposing, our key term mode is intended to categorise and modularise physical behaviours. Crucially, the modes of behaviour overlap which means there are transitions between modes. Theoretically, the modes are sets of states of the system; they may have mathematical models but \textit{a priori} they are not associated with any software specification. For each mode $\alpha$ we have a set of data $X_\alpha$ necessary to describe the system, rules for its evolution and (subtle) notions of modes overlapping  To tackle the transition from a set of modes we construct the \textit{nerve}, which is simply the combinatorial representation of the geometry of the system, based on considering the overlapping of the subsets defining the modes.

A hybrid automaton may be nondeterministic, meaning that more than one arrow from a state may have its guard condition satisfied. For our modes, however, the structure of the nerve and the idea that all information has errors forces nondeterminism. In fact, where three modes have a simultaneous intersection, we expect a whole continuum of possibilities ranging between two arrows each being an `obvious' one to take. This nondeterminism is linked to the geometry of the state space. From a topological point of view, it corresponds to homotopic paths through state space. The continuum of possibilities in evaluating the arrows is encapsulated into a \textit{mode evaluation} function, which lies at the heart of our approach. 

It is instructive to ask a question on making a change of discrete state allowed by guard clause, or in our terms a mode transition. Is this change determined by the behaviour of the system, or is it within the software's control (ultimately the software designer's control) to decide what to do? In terms of the popular heating system example, is it the laws of physics which dictate that the system changes, or does a user decide that the room is too hot? We separate out these possibilities to clarify our thinking and to emphasise that it is really the real world system which is in control -- the software can only respond to data about the physical behaviour; the distinction may enhance the reusability of the software. Every mode comes with an externally imposed set of orders. The mode evaluation function, which amongst other jobs evaluates how well the current mode is performing, does so by reference to the orders. Some discrete transitions are forced by the system, and others are chosen with reference to the orders. 

Finally, we should refer to the Zeno condition, which is that the system should not try to execute infinitely many transitions in finite time. From our point of view, this can't happen, as the transitions have definite finite durations -- in fact estimates of the transition times would be required. To explain why this works, part of the job of mode evaluation is to estimate whether a potential mode to transfer to has a long enough expected duration to evaluate its own performance (including real world measurements) and its possible transfers out -- do not transfer into a mode unless that mode can be transferred out of in an orderly fashion. Of course, because of errors in measurements, unexpected behaviour of the system, or just bad programming, that may not be possible, and we may be left with various sub-optimal choices, a matter which we shall discuss -- that is our version of the Zeno problem.

\subsection{Structure of the paper}\label{StructurePaper}
The structure of the paper is this. In  Section \ref{Principles}, we formulate some general principles for analogue-digital systems to guide our thinking and mathematical formulations. In Section \ref{Physical_modes}, we attempt to clarify some intuitions about modes of operation, and transitions between modes, by introducing an example of a racing car under the control of software. 

In Section \ref{AD-systems}, we give the general definition of an analogue-digital system with modes, and introduce a simple software architecture to organise our discussion.  The next tasks are to show how to formalise modes and mode transitions, i.e., 

(i) to define modes; 

(ii) to evaluate their fitness for purpose; and 

(iii) to change from one mode to another. 

\noindent In Sections \ref{Mode_implementation} we address (i) and in Section \ref{Mode_evaluation} we address (ii).  In Section \ref{Mode_transition}, the transition theory is developed to address (iii).

Next, we turn to formal examples to illustrate the theory. In Section \ref{cadsbqil}, we give a general mathematical example that characterises many classical systems based on manifolds and geodesics. These illustrate modes in the cases where one can usefully imagine the presence of some global state space based upon combining the state spaces of many modes.  In Section \ref{chicane} we describe an example based on the informal car example in Section \ref{Physical_modes}.  In Section \ref{solar}, we describe an example based on the orbits of planets.

Finally, in Section \ref{Concluding_remarks}, we return to hybrid systems and touch on topics arising from the study: networks of analogue-digital systems; multi-scale analysis, when we look at the system in finer or coarser detail; and security. 

We thank Felix Costa (Lisbon) for many enjoyable and influential conversations on analogue-digital systems; this paper has grown out of our collaboration on  \cite{Oracles1,Oracles2,Wheatstone,Axiomatising,ImapctModels,ThreeExperiments}.

\section{Some principles for analogue-digital system specification}\label{Principles}

We propose to make a theory that can speak about three distinct components of a complex analogue-digital system:  (i) the physical system as it is in reality; (ii) the modes of behaviour of the physical system; and (iii) the algorithms and software that employ the modes to control the physical system. Let us clarify these three distinctions by adopting five more working principles.
\newline
\newline
\noindent \textbf{Principle 1: Physical uniqueness.} \textit{A physical system, process or environment is based on equipment made from particular physical components and whose behaviour is a continuous physical phenomenon that is {\em unique} to that equipment in space and time.}
\newline
\newline
\noindent \textbf{Principle 2: Physical observation.} \textit{A physical system, process or environment does what it does. To {\em observe and measure} a system is to abstract aspects of its behaviour in a way that is specific to the system. The observations and measurements are the basis of controlling the physical situation. They are the basis for interfacing algorithms and physical behaviour, and for classifying the behaviour of the system into {\em modes}.}
\newline
\newline
\noindent \textbf{Principle 3: Physical specification.} \textit{To {\em model} a mode of behaviour of a system by a dynamical system is to specify an aspect of the physical system in an abstract and general way; the model specifies a {\em class} of particular physical systems having that mode of behaviour. In addition to providing understanding of how the physical system might behave, dynamical systems provide a specification of a general data type interface for a mode.}
\newline
\newline
The dynamical systems are central to the study of the software. The models of a physical system may be many and varied; they are shaped by different choices of physical insights, spatial and temporal scales, and computational constraints, etc. There are many ways of modelling analytically (such as partial differential equations, ordinary differential equations) and computationally (such as finite elements, neural networks, lattices, cellular automata.)
\newline
\newline
\noindent \textbf{Principle 4: Modes and Models.} \textit{A complex system will have many modes. {\em Ideally}, each mode will have a sound model. A {\em portfolio of modes and models}, with overlapping domains of application, is needed to cover the system's behaviour.}
\newline
\newline
Even if each mode of behaviour has its own individual model with a state space, the models can be disparate, and the system does \textit{not} have a unified model or state space. 

Recalling the  \textit{Principle 0: Robustness} from the Introduction, the aspiration is that all physical eventualities are covered:
\newline
\newline
\noindent \textbf{Principle 5: Mode Transitions.} \textit{A methodology for managing a portfolio of models is needed that has methods for 
\newline
(i) evaluating the quality of the models in the portfolio;
\newline
(ii) passing from one model to another; and
\newline
(iii) dealing with exceptional situations when {\em no model in the portfolio applies}.}
\newline
\newline
The fact that the modes are physical processes observed by measurements that are liable to errors is fundamental to our investigation.

\section{Physical modes and why they may change}\label{Physical_modes}

We begin by exploring the need for modes of operation and transitions between modes, informally, by means of a thought experiment on autonomous cars. 

\subsection{Example of a system with many modes: a robotic racing car} \label{carrace}
Consider an autonomous car racing championship. To create the software to govern the racing car, the programmers are given:
\newline
\newline
\noindent \textit{Observables.} Specifications of the data available from the sensors and actuators, and of their behaviour; for example, `` the maximum acceleration is $10m/sec^2$"; ``the time delay on the speedometer reading is $0.5 sec$"; ...). 
\newline
\newline
\noindent \textit{Objectives.} Driving tactics for the race; for example,  ``minimise the lap time"; ``don't get too close to the other cars or the edge of the track"; ``beware reckless car A"; ...). 
\newline
\newline
First, the programmers must separate concerns to simplify the problem. Immediately, the behaviour of the physical system is split into modes, chosen to simplify the mathematical modelling of observables of the behaviour and the specification of driving objectives.  For our racing car, we might manage with five modes as follows:

\smallskip
\textit{Mode $\alpha$: The car is on the track and some distance from the nearest cars.} 

\smallskip\noindent The state of any other car is reduced simply to the position of a point on the track, as it is not particularly relevant. The important objectives are maximising acceleration and taking corners at the fastest safe speed; and perhaps some longer term tactics, such as when to refuel or do a wheel change. 

\smallskip
It would be nice to stay in mode $\alpha$ all the time, but this is unlikely as at some point the car will meet rival cars. This meeting can be determined from the picture available from the state space of mode  $\alpha$ by looking at the distance to the nearest cars. When this distance becomes smaller than a certain value, the car will need to change behaviour. This change in separation ought to be expected and predicted by a model; but it could also be unexpected (e.g., the car hits a pool of oil and skids).

\smallskip
\textit{Mode $\beta$: The car is on the track but close to other cars.} 

\smallskip\noindent The objectives are to avoid collision, and to overtake if possible. Now more state variables are needed to handle the nearby cars. For example, to positions of the other cars, their velocities, accelerations, and sizes may be added. Also, perhaps, their identities and some intelligence may be added (e.g., ``car A always tried to block us when we tried to overtake, car B did not"). 

\smallskip
At a particular combination of positions and velocities of nearby cars, a collision may be flagged up as a real danger. 

\smallskip
\textit{Mode $\gamma$: The car is on the track but on a crash course with another car.} 

\smallskip\noindent The objective becomes not only to see if it is possible to avoid a collision, but to mitigate the effects if it cannot be avoided. At this point the actual size and shape of the cars may be important, if (say) a wheel on wheel impact was likely to be more dangerous than a side on side impact. 

\smallskip
These are three simple modes that make an initial design of a system; maybe three modes could win a race, but there are gaps. Suppose the car comes to a complete stop ending up off the track, undamaged on a grass verge: the programmers need to bolt on another mode: 

\smallskip
\textit{Mode $\delta$: The car is off the track.}

\smallskip \noindent 
The objective is to rejoin the track as soon as possible, but to stay away from cars on the track. 

Experience generates further problems that must be considered. Suppose the instrumentation develops a fault and core data is untrustworthy. As the software was not designed to cope with (say) incorrect readings from the speedometer, a crash could result. To fix this the programmers need to add another mode:

\smallskip
\textit{Mode $\epsilon$: The car is on the track but the instrumentation is faulty.}

\smallskip\noindent The objective is to slow down, avoid other cars and to make way to the grass verge and stop.
To add this mode, the programmers need to carry out some modifications to the previous ones. These modifications are not major, they just need to say when the error mode $\epsilon$ should be entered. For example, a comparison of the speedometer reading can be made with the changes of positions in time of objects on the side of the track. If there is an obvious disagreement, go to mode $\epsilon$. This modification is for modes $\alpha$ and $\beta$: if we are in mode $\gamma$, we already have more urgent problems.

A useful distinction is to declare the modes to be \textit{normal} or \textit{exceptional}, for example: modes $\alpha$, $\beta$, and $\gamma$ are normal and $\delta$ and $\epsilon$ are exceptional.

\subsection{A modular approach to modelling complicated physical systems}

From the point of view of designing an analogue-digital system, such as a driverless racing car, the modes represent a form of physical modularisation, determined by the science of the physical domain and, possibly, exceptional circumstances. Consider two aspects of physical modelling: modularity and accuracy.

\smallskip
\noindent \textbf{Modularity.} The benefits of modularity are well known, but here modularisation is a necessity. Physical systems possess all sorts of modes, not least modes for startup, normal running, exceptional situations, and shutdown. Mathematical models of physical behaviour are approximations with limited ranges of validity. In complex environments and systems, physical conditions change, and the range of validity of a model can easily be stressed and broken. Thus, it is to be \textit{expected} that \textit{a change of physical conditions requires a change of mathematical model}. In this way complex systems acquire different modes, where the entire description of the system may change, and a completely new set of state variables and a new type of model is used. 

The modes constitute a modular approach to understanding physically the system so that 

\noindent \textit{Specification.} We can track of properties that characterise a mode and trigger need for a change.
\newline
\noindent \textit{Implementation.} When a mode is modified in a minor way, or more special cases added, we can isolate and limit the changes  -- and consequent errors -- in software. 
\newline
\noindent \textit{Efficiency.} We can detect if a problem can be solved faster with parallel processing.

The models for the modes may be of disparate kinds: they may be analytical models based on PDEs or coupled systems of ODEs; or they may be algorithmic models based on numerical approximation methods, neural networks, coupled map lattices, and cellular automata. In a complex system one can expect several kinds to be present. 

There are methods of combining algorithmic models in different domains because algorithmic models are discrete space, discrete time models and can be unified in a theory of synchronous concurrent algorithms \cite{TTZ}; for example, in the case of whole heart modelling, see \cite{PTHa,PTHb,CCHT}.

\smallskip
\noindent \textbf{Accuracy.} In a simple physical system, with easy access to measurements, it may be reasonable to assume that the software has instant access to a comprehensive specification of the physical system at any time. However, in practice, the system may be complicated enough to require some factors to be estimated. It would be helpful to have a modular approach to modelling the system that can cope with factors such as:
\newline
\noindent \textit{Delay.} Measurements take time to perform, or be available only to a limited accuracy. There may be a time-lag in obtaining some information that requires action by the software.
\newline
\noindent \textit{Decay.} Sensors, actuators and communication links are all prone to degradation (or even complete failure) and replacement will take time, if it is possible at all. If the performance of a component has degraded, the software may have to spot this from inconsistencies in the data, and take appropriate action.\footnote{The consequences of unreliable approximation to reality can be catastrophic, for example, on an airliner crash in the Atlantic following bad weather:
{ \quote 
\textit{Temporary inconsistency between the measured airspeeds, likely following the
obstruction of the Pitot probes by ice crystals that led in particular to autopilot
disconnection and a reconfiguration to alternate law,} -- Final Report
On the accident on 1st June 2009, Bureau d'Enqu\^etes et d'Analyses
pour la s\'ecurit\'e de l'aviation civile
\endquote}} 
\newline
\noindent  \textit{History.} If there is no recent data to base a decision on, historical data may be needed to make a best guess.

\section{Analogue-digital systems}\label{AD-systems}

We define a general analogue-digital system without modes and then generalise the definition by introducing modes.

\subsection{What is a general analogue-digital system?}\label{AD-system}

The basic architecture of an analogue-digital system is depicted in Figure 2, in which arrows are used to indicate the direction of flow of information.

\unitlength 0.5 mm

\begin{picture}(240,90)(10,0)
\linethickness{0.3mm}
\put(-2,80){\line(1,0){52}}
\put(-2,20){\line(0,1){60}}
\put(50,20){\line(0,1){60}}
\put(-2,20){\line(1,0){52}}
\linethickness{0.3mm}
\put(100,80){\line(1,0){40}}
\put(100,20){\line(0,1){60}}
\put(140,20){\line(0,1){60}}
\put(100,20){\line(1,0){40}}
\linethickness{0.3mm}
\put(190,80){\line(1,0){50}}
\put(190,20){\line(0,1){60}}
\put(240,20){\line(0,1){60}}
\put(190,20){\line(1,0){50}}
\put(215,70){\makebox(0,0)[cc]{data}}
\put(215,52.5){\makebox(0,0)[cc]{algorithm}}
\put(215,35){\makebox(0,0)[cc]{\textit{objectives}}}

\put(24,70){\makebox(0,0)[cc]{physical system}}
\put(24,35){\makebox(0,0)[cc]{\textit{laws or theories}}}

\put(120,70){\makebox(0,0)[cc]{interface}}
\put(120,35){\makebox(0,0)[cc]{\textit{axioms}}}

\linethickness{0.3mm}
\put(55,60){\line(1,0){40}}
\put(95,60){\vector(1,0){0.12}}
\linethickness{0.3mm}
\put(55,50){\line(1,0){40}}
\put(55,50){\vector(-1,0){0.12}}
\put(75,66){\makebox(0,0)[cc]{sensor bus}}

\put(75,45){\makebox(0,0)[cc]{actuator bus}}

\linethickness{0.3mm}
\put(145,60){\line(1,0){40}}
\put(145,50){\line(1,0){40}}
\put(185,60){\vector(1,0){0.12}}
\put(145,50){\vector(-1,0){0.12}}

\put(163,45){\makebox(0,0)[cc]{queries}}
\put(163,65){\makebox(0,0)[cc]{responses}}

\put(125,6){\makebox(0,0)[cc]{\textbf{Figure 2:} A analogue-digital system}}

\end{picture}

The \textit{physical system} is a process, system or environment that we wish to monitor or control. Suppose there is a fixed system of sensors for measurements, and actuators for control, linked to the physical system; so we should think of information flow along a \textit{sensor bus} (e.g., from thermometers) and an \textit{actuator bus} (e.g., to motors). 

The \textit{interface} is what translates between the sensors and actuators and the decision making software. The interface can take many forms; in a very simple case of the direct control of a motor by a microprocessor, it would be little more than some digital-to-analogue (DA) and analogue-to-digital (AD) converters. The interface may need to manage error margins and timing delays. 

A \textit{query} is a request from the decision making software to the interface and a \textit{response} is a message going the other way. In the motor and microprocessor example, the query is simply a binary number interpreted by the interface's DA-converter as a voltage applied to a motor, and there is no response. The response to a query normally takes time and it is possible that the response is timed out.

Finally, there is the \textit{algorithm} that formalises the decision making that attempts to monitor or control the physical system. The data must be considered separately, as it is the sole basis by which the algorithm knows what is happening in the physical system, it forms the algorithm's picture of physical reality. The data must be updated, either periodically by standing orders in the interface, or when the algorithm sends a query to update items. 

The remaining components of Figure 2, mentioned in italics, are the basis of \textit{specifications}. The \textit{laws or theories} refer to mathematical models of the physical system under consideration, e.g.,\ Newton's laws of motion or characteristics of electronic devices. Using these laws, and the sensors and actuators, an \textit{axiomatic specification} of the interface is written.\footnote{For example, such axioms might be:

``System variable $x$ can be measured to accuracy $\epsilon$ in time $t$."

``System variable $y$ obeys the differential equation to within error $\epsilon$."

``System variable $z$ always remains in the interval $[-\pi,+\pi]$."}

The axiomatic description of the interface is used to write an algorithm to satisfy the list of \textit{objectives}.\footnote{For example, such an objective might be:
``System variable $x$ should always satisfy $x\le 30$."} For given initial configurations, if it can be proved that the algorithm suffices to guarantee the objectives then the algorithm can be said to be \textit{verified assuming the validity of the axioms}. 
%If it is not possible to guarantee that the objectives are always satisfied, a probabilistic approach may be used. 

\subsection{Physical Oracles}\label{oracles}

The central idea is that \textit{the physical system or environment is an oracle to the algorithm}. The algorithm requests and receives data about the physical environment or system to use in its processing. The interface between the analogue and digital is of central theoretical interest. Thus, we have in mind a \textit{physical oracle}, in which the computer may \textit{query} the interface, and receive messages back. The queries may be of various forms, e.g.,\ a question for a temperature sensor, or an instruction for a motor to be turned on. The algorithm need not know anything about the detailed operation of sensors and actuators, that is all abstracted in the interface.\footnote{The interfaces are far more complicated than the traditional oracles of computability theory. There, since Emil Post's development of computability theory, a query asks if a datum is in a set, and the response is yes or no and takes one time step.}

The idea of an oracle is very general and is not confined to computing systems.  For example, a complicated socio-technical example would be a laboratory service in a hospital. Instructions are received from doctors to perform tests, and the results are sent back to the doctors. There is a division of labour: the laboratory staff do not need to know about diagnosing disease, and the doctor does not need to know how to perform the tests. The different times taken for tests, the queue of requests and the capacity of the lab mean that the lab staff have to make scheduling decisions. If a request is urgent, they may have to decide between a quick but less reliable method and a slower more reliable test, of course notifying the doctor of the possible errors concerned. In this example, a query by a doctor might be ``measure levels of X in patient Y by 09.00 tomorrow'', and a response might be ``level is Z, less accurate method used due to urgency of test''. The choice of test is a choice of mode. The idea of oracle can be applied similarly to any service in which time is a prominent factor.  

\subsection{An analogue-digital system with modes of operation}\label{AD-system_with_modes}
Consider an analogue-digital system whose physical behaviour can be split into \textit{modes}. Starting with Figure 2, we imagine an architecture for three modes $\alpha, \beta, \gamma$ depicted in Figure 3:

\unitlength 0.5 mm
\begin{picture}(240,112)(10,-10)
\linethickness{0.3mm}
\put(90,90){\line(1,0){40}}
\put(90,70){\line(0,1){20}}
\put(130,70){\line(0,1){20}}
\put(90,70){\line(1,0){40}}
\linethickness{0.3mm}
\put(90,60){\line(1,0){40}}
\put(90,40){\line(0,1){20}}
\put(130,40){\line(0,1){20}}
\put(90,40){\line(1,0){40}}
\linethickness{0.3mm}
\put(90,30){\line(1,0){40}}
\put(90,10){\line(0,1){20}}
\put(130,10){\line(0,1){20}}
\put(90,10){\line(1,0){40}}
\linethickness{0.3mm}
\put(0,80){\line(1,0){40}}
\put(0,20){\line(0,1){60}}
\put(40,20){\line(0,1){60}}
\put(0,20){\line(1,0){40}}
\linethickness{0.3mm}
\put(40,65){\line(1,0){20}}
\linethickness{0.3mm}
\put(60,25){\line(0,1){60}}
\linethickness{0.3mm}
\put(60,85){\line(1,0){25}}
\put(85,85){\vector(1,0){0.12}}
\linethickness{0.3mm}
\put(60,55){\line(1,0){25}}
\put(85,55){\vector(1,0){0.12}}
\linethickness{0.3mm}
\put(60,25){\line(1,0){25}}
\put(85,25){\vector(1,0){0.12}}
\linethickness{0.3mm}
\put(45,45){\line(1,0){10}}
\put(45,45){\vector(-1,0){0.12}}
\linethickness{0.3mm}
\put(65,45){\line(1,0){5}}
\linethickness{0.3mm}
\put(70,60){\line(0,1){15}}
\linethickness{0.3mm}
\put(70,30){\line(0,1){20}}
\linethickness{0.3mm}
\put(70,15){\line(0,1){5}}
\linethickness{0.3mm}
\put(70,75){\line(1,0){20}}
\linethickness{0.3mm}
\put(70,45){\line(1,0){20}}
\linethickness{0.3mm}
\put(70,15){\line(1,0){20}}
\linethickness{0.3mm}
\put(180,90){\line(1,0){50}}
\put(180,70){\line(0,1){20}}
\put(230,70){\line(0,1){20}}
\put(180,70){\line(1,0){50}}
\linethickness{0.3mm}
\put(180,60){\line(1,0){50}}
\put(180,40){\line(0,1){20}}
\put(230,40){\line(0,1){20}}
\put(180,40){\line(1,0){50}}
\linethickness{0.3mm}
\put(180,30){\line(1,0){50}}
\put(180,10){\line(0,1){20}}
\put(230,10){\line(0,1){20}}
\put(180,10){\line(1,0){50}}
\linethickness{0.5mm}
\put(170,95){\line(1,0){70}}
\put(170,5){\line(0,1){90}}
\put(240,5){\line(0,1){90}}
\put(170,5){\line(1,0){70}}
\linethickness{0.3mm}
\put(135,85){\line(1,0){40}}
\put(175,85){\vector(1,0){0.12}}
\linethickness{0.3mm}
\put(135,55){\line(1,0){40}}
\put(175,55){\vector(1,0){0.12}}
\linethickness{0.3mm}
\put(135,25){\line(1,0){40}}
\put(175,25){\vector(1,0){0.12}}
\linethickness{0.5mm}
\put(80,95){\line(1,0){60}}
\put(80,5){\line(0,1){90}}
\put(140,5){\line(0,1){90}}
\put(80,5){\line(1,0){60}}
\put(20,61){\makebox(0,0)[cc]{physical}}

\put(20,50){\makebox(0,0)[cc]{system}}

\put(109,81){\makebox(0,0)[cc]{\textsf{interface}${}_\alpha$}}

\put(204,76){\makebox(0,0)[cc]{algorithm${}_\alpha$}}
\put(204,85){\makebox(0,0)[cc]{data${}_\alpha$}}

\put(109,51){\makebox(0,0)[cc]{\textsf{interface}${}_\beta$}}

\put(204,46){\makebox(0,0)[cc]{algorithm${}_\beta$}}
\put(204,55){\makebox(0,0)[cc]{data${}_\beta$}}

\put(109,21){\makebox(0,0)[cc]{\textsf{interface}${}_\gamma$}}

\put(204,16){\makebox(0,0)[cc]{algorithm${}_\gamma$}}
\put(204,25){\makebox(0,0)[cc]{data${}_\gamma$}}

\put(58,91){\makebox(0,0)[cc]{sensor bus}}

\put(56,11){\makebox(0,0)[cc]{actuator bus}}

\linethickness{0.3mm}
\put(135,75){\line(1,0){40}}
\put(135,75){\vector(-1,0){0.12}}
\linethickness{0.3mm}
\put(135,45){\line(1,0){40}}
\put(135,45){\vector(-1,0){0.12}}
\linethickness{0.3mm}
\put(135,15){\line(1,0){40}}
\put(135,15){\vector(-1,0){0.12}}

\put(123,-7){\makebox(0,0)[cc]{\textbf{Figure 3:} A three mode analogue-digital system}}

\end{picture}

\smallskip
\smallskip
Domain scientists and engineers must classify the behaviour of the physical situation to (i) define the modes and (ii) create the list of objectives for each mode.  We suppose each mode $\alpha\in \mathcal{M}$ has a list of state variables, a state space and trajectories that represent behaviour in time. These may be derived from a model that is a reasonable mathematical description of the physical system in that mode, or from some ad hoc rules intended to cope with unusual situations. In each mode, appropriate approximations may be made to keep the description of the mode manageable. 

Thus, in the analogue-digital system a component \textsf{state}${}_\alpha$ managing the system in mode $\alpha$ will contain the state variables for mode $\alpha$, and probably more, such as estimates of the errors of the state variables, and some of the known properties of the system. 

In addition to coding the modes $\alpha\in \mathcal{M}$, there must also be conditions that have to be monitored and that, if satisfied, cause a change from mode $\alpha$ to some new mode $\beta$. 

A \textit{normal mode} is one in which the system is functioning well in delivering its basic objectives. An \textit{exceptional mode} is essentially an error recovery situation and is not considered a desirable operation. The aim is the solve the 
following:
\newline
\newline
\textbf{Validation Problem.} In designing an analogue-digital system with many modes we wish to validate -- by testing and verification -- that, relative to assumptions about the physical components,

1. There are sufficiently many modes to cover all cases, including exceptional modes; and 

2. Given a subset of initial states, the behaviour of the system over a given time interval can be guaranteed to stay within the normal modes.

\section{Designing a mode implementation}\label{Mode_implementation}

\subsection{The architecture of a multimodal analogue-digital system}\label{Architecture_AD-system}
The architecture is designed to gather information about physical behaviour; monitor and evaluate the behaviour as a mode; and supervise a mode transition if needed. The components needed to make the software are summarised in the architecture depicted in Figure 4; compared with Figure 3, the physical system is omitted and, for simplicity, now only two modes $\alpha$ and $\beta$, and one transition from mode $\alpha$ to mode $\beta$, are shown. The next three sections will be devoted to explaining the principles behind this choice of components and architecture.

\unitlength 0.5 mm
\begin{picture}(265,125)(28,-6)
\linethickness{0.3mm}
\put(20,105){\line(1,0){40}}
\put(20,75){\line(0,1){30}}
\put(60,75){\line(0,1){30}}
\put(20,75){\line(1,0){40}}
\linethickness{0.3mm}
\put(20,50){\line(1,0){40}}
\put(20,20){\line(0,1){30}}
\put(60,20){\line(0,1){30}}
\put(20,20){\line(1,0){40}}
\linethickness{0.3mm}
\put(90,107.5){\line(1,0){40}}
\put(90,92.5){\line(0,1){15}}
\put(130,92.5){\line(0,1){15}}
\put(90,92.5){\line(1,0){40}}
\linethickness{0.3mm}
\put(90,77.5){\line(1,0){40}}
\put(90,62.5){\line(0,1){15}}
\put(130,62.5){\line(0,1){15}}
\put(90,62.5){\line(1,0){40}}
\linethickness{0.3mm}
\put(90,47.5){\line(1,0){40}}
\put(90,32.5){\line(0,1){15}}
\put(130,32.5){\line(0,1){15}}
\put(90,32.5){\line(1,0){40}}
\linethickness{0.3mm}
\put(90,17.5){\line(1,0){40}}
\put(90,2.5){\line(0,1){15}}
\put(130,2.5){\line(0,1){15}}
\put(90,2.5){\line(1,0){40}}
\linethickness{0.3mm}
\put(155,85){\line(1,0){40}}
\put(155,65){\line(0,1){20}}
\put(195,65){\line(0,1){20}}
\put(155,65){\line(1,0){40}}
\linethickness{0.3mm}
\put(155,25){\line(1,0){40}}
\put(155,5){\line(0,1){20}}
\put(195,5){\line(0,1){20}}
\put(155,5){\line(1,0){40}}
\linethickness{0.3mm}
\put(220,105){\line(1,0){40}}
\put(220,5){\line(0,1){100}}
\put(260,5){\line(0,1){100}}
\put(220,5){\line(1,0){40}}
\linethickness{0.3mm}
\put(20,90){\line(1,0){40}}
\linethickness{0.3mm}
\put(20,35){\line(1,0){40}}

%%%%%
\put(40,97){\makebox(0,0)[cc]{\textsf{query}${}_\alpha$}}

\put(40,67){\makebox(0,0)[cc]{\textsf{response}${}_\alpha$}}

\put(40,83){\makebox(0,0)[cc]{\textsf{interface}${}_\alpha$}}

\put(110,70){\makebox(0,0)[cc]{\textsf{state}${}_\alpha$}}

\put(110,100){\makebox(0,0)[cc]{\textsf{control}${}_\alpha$}}

\put(175,100){\makebox(0,0)[cc]{\textsf{orders}${}_\alpha$}}

\put(175,75){\makebox(0,0)[cc]{\textsf{monitor}${}_\alpha$}}

\put(40,41){\makebox(0,0)[cc]{\textsf{query}${}_\beta$}}

\put(40,12){\makebox(0,0)[cc]{\textsf{response}${}_\beta$}}

%%%%%%

\put(40,27){\makebox(0,0)[cc]{\textsf{interface}${}_\beta$}}

\put(110,10){\makebox(0,0)[cc]{\textsf{state}${}_\beta$}}

\put(110,40){\makebox(0,0)[cc]{\textsf{control}${}_\beta$}}

\put(175,40){\makebox(0,0)[cc]{\textsf{orders}${}_\beta$}}

\put(175,15){\makebox(0,0)[cc]{\textsf{monitor}${}_\beta$}}

\put(160,54){\makebox(0,0)[cc]{$\tau_{\beta\alpha}$}}

\put(240,60){\makebox(0,0)[cc]{\textsf{supervisor}}}

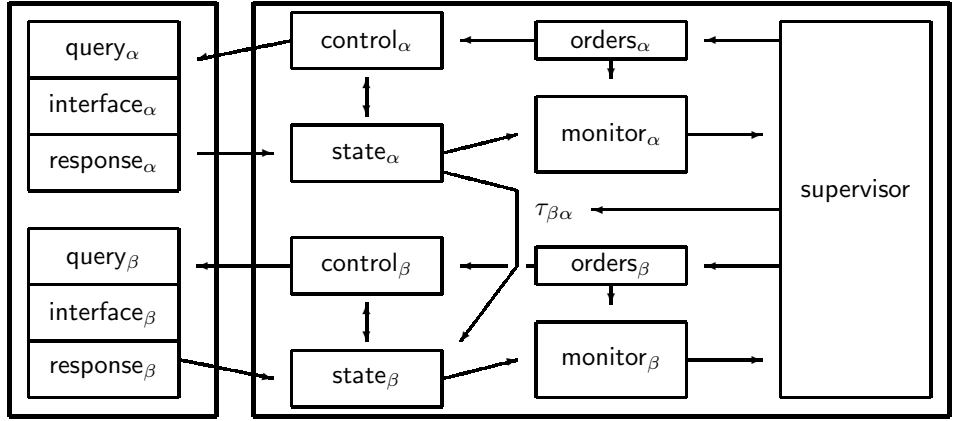
\put(135,-14){\makebox(0,0)[cc]{\textbf{Figure 4:} Details of a two mode analogue-digital system}}

%%%%%

\linethickness{0.3mm}
\multiput(65,95)(0.6,0.12){42}{\line(1,0){0.6}}
\put(65,95){\vector(-4,-1){0.12}}
\linethickness{0.3mm}
\put(65,40){\line(1,0){25}}
\put(65,40){\vector(-1,0){0.12}}
\linethickness{0.3mm}
\put(65,70){\line(1,0){20}}
\put(85,70){\vector(1,0){0.12}}
\linethickness{0.3mm}
\multiput(60,15)(0.6,-0.12){42}{\line(1,0){0.6}}
\put(85,10){\vector(4,-1){0.12}}
\linethickness{0.3mm}
\multiput(130,70)(0.48,0.12){42}{\line(1,0){0.48}}
\put(150,75){\vector(4,1){0.12}}
\linethickness{0.3mm}
\multiput(130,10)(0.48,0.12){42}{\line(1,0){0.48}}
\put(150,15){\vector(4,1){0.12}}
\linethickness{0.3mm}
\put(195,75){\line(1,0){20}}
\put(215,75){\vector(1,0){0.12}}
\linethickness{0.3mm}
\put(195,15){\line(1,0){20}}
\put(215,15){\vector(1,0){0.12}}
\linethickness{0.3mm}
\multiput(130,65)(0.48,-0.12){42}{\line(1,0){0.48}}
\linethickness{0.3mm}
\put(150,40){\line(0,1){20}}
\linethickness{0.3mm}
\multiput(135,20)(0.12,0.16){125}{\line(0,1){0.16}}
\put(135,20){\vector(-3,-4){0.12}}

\linethickness{0.3mm}
\put(170,55){\line(1,0){50}}
\put(170,55){\vector(-1,0){0.12}}
\linethickness{0.5mm}
\put(80,110){\line(1,0){185}}
\put(80,0){\line(0,1){110}}
\put(265,0){\line(0,1){110}}
\put(80,0){\line(1,0){185}}
\linethickness{0.5mm}
\put(15,110){\line(1,0){55}}
\put(15,0){\line(0,1){110}}
\put(70,0){\line(0,1){110}}
\put(15,0){\line(1,0){55}}
\linethickness{0.3mm}
\put(110,80){\line(0,1){10}}
\put(110,90){\vector(0,1){0.12}}
\put(110,80){\vector(0,-1){0.12}}
\linethickness{0.3mm}
\put(110,20){\line(0,1){10}}
\put(110,30){\vector(0,1){0.12}}
\put(110,20){\vector(0,-1){0.12}}
\linethickness{0.3mm}
\put(155,105){\line(1,0){40}}
\put(155,95){\line(0,1){10}}
\put(195,95){\line(0,1){10}}
\put(155,95){\line(1,0){40}}
\linethickness{0.3mm}
\put(155,45){\line(1,0){40}}
\put(155,35){\line(0,1){10}}
\put(195,35){\line(0,1){10}}
\put(155,35){\line(1,0){40}}
\linethickness{0.3mm}
\put(200,100){\line(1,0){20}}
\put(200,100){\vector(-1,0){0.12}}
\linethickness{0.3mm}
\put(200,40){\line(1,0){20}}
\put(200,40){\vector(-1,0){0.12}}
\linethickness{0.3mm}
\put(135,100){\line(1,0){20}}
\put(135,100){\vector(-1,0){0.12}}
\linethickness{0.3mm}
\put(135,40){\line(1,0){12.5}}
\put(135,40){\vector(-1,0){0.12}}
\linethickness{0.3mm}
\put(175,90){\line(0,1){5}}
\put(175,90){\vector(0,-1){0.12}}
\linethickness{0.3mm}
\put(175,30){\line(0,1){5}}
\put(175,30){\vector(0,-1){0.12}}
\linethickness{0.3mm}
\put(152.5,40){\line(1,0){2.5}}
\linethickness{0.3mm}
\put(20,5){\line(0,1){15}}
\linethickness{0.3mm}
\put(20,5){\line(1,0){40}}
\linethickness{0.3mm}
\put(60,5){\line(0,1){15}}
\linethickness{0.3mm}
\put(20,60){\line(0,1){15}}
\linethickness{0.3mm}
\put(20,60){\line(1,0){40}}
\linethickness{0.3mm}
\put(60,60){\line(0,1){15}}
\end{picture}

\subsection{The architecture of a mode}\label{mode_architecture}
For each mode $\alpha\in \mathcal{M}$ we have the following components:

\smallskip
\noindent\textbf{\textsf{state${}_\alpha$}}: 
The data in this component constitutes mode $\alpha$'s picture of the physical system, i.e., the state space ${\mathrm{State}}_\alpha$. It can be broadly divided into \textit{history}, \textit{observations}, \textit{deductions} or \textit{estimations}. Also system variables are likely to be accompanied by estimates of their errors, and the time of last measurement. Much of this data may be unused by \textsf{control}${}_\alpha$, but more of it is likely to be used by \textsf{monitor}${}_\alpha$ (more on this later). 
\smallskip
\noindent \textbf{\textsf{interface}${}_\alpha$}: This is the only component of the program that communicates with the physical system. Its input to the interface from the rest of the program are queries from a set ${\mathrm{Queries}}_\alpha$, and lie in its ${\mathrm{query list}}_\alpha$, being the list of tasks to be performed with the physical system. Queries are requests to make observations, and controlling instructions. The outputs from the interface to the rest of the program are the responses from a set ${\mathrm{Responses}}_\alpha$ and are passed to the component \textsf{state}${}_\alpha$.  Responses may be values it observes, and answers to questions. The interface has a default list of tasks designed to keep the states in \textsf{state}${}_\alpha$ up to date. The tasks may be assigned various priorities, e.g.,\ the tuples

$\big<\mathrm{single\ action},\ \mathrm{urgent\ priority},\ \mathrm{apply\ brakes}\big>$

$\big<\mathrm{repeat\ hourly},\ \mathrm{measure\ temperature},\ \mathrm{accuracy\ 1}^\circ\big>$

$\big<\mathrm{single\ action},\ \mathrm{standard\ priority},\ \mathrm{measure\ temperature},\ \mathrm{accuracy \ 1}^\circ,$

$\mathrm{if\ previous\ measurement\ older\ than\ 1\ minute}\big>$

\noindent Here we see that some tasks may be more urgent than others, while some may be instructions to be repeated. They may come with conditions for the measurements to be made, or with additional information, such as a required accuracy. 

 Why is this rather indirect method used for communication? One answer is that the controlling algorithm is really not interested in the details of a particular measuring device. Should a sensor wear out and be replaced by one from a different manufacturer, only the interface would have to be updated.

\smallskip
\noindent\textbf{\textsf{control}${}_\alpha$}: This component of mode $\alpha$ actually makes decisions about controlling the physical system. It reads data from components \textsf{state}${}_\alpha$ and \textsf{orders}${}_\alpha$, and writes to \textsf{state}${}_\alpha$ (typically \textit{estimations}) and the  \textit{query list} for \textsf{interface}${}_\alpha$. It might have some ability to respond to observed problems (e.g.,\ by increasing the frequency of making some observations).

\smallskip
\noindent \textbf{\textsf{orders}${}_\alpha$}: While the response of \textsf{control}${}_\alpha$ may be largely determined by the code, there may be some flexibility in its behaviour. This could be viewed as setting the overall strategy for mode $\alpha$.  Let  \textsf{orders}${}_\alpha$ be the set of possible instructions for mode $\alpha$. A simple example would be temperature control for a building, where the inhabitants could set the desired temperature range, and this range would be put in as an instuction in \textsf{orders}${}_\alpha$. 

\smallskip
\noindent 
\textbf{Definition.}  A \textit{physical mode}  $\alpha$ is specified by a set ${\mathrm{State}}_\alpha$ of states whose role is 

\noindent (i) \textit{Monitoring}: to observe and act upon the behaviour of the system; and 

\noindent (ii) \textit{Reflexivity}: to evaluate its own performance in faithfully representing that behaviour.  

\noindent

Various functions on ${\mathrm{State}}_\alpha$ arise in specifying examples. For example, in monitoring, one expects functions to represent the generation of queries and the collection of responses:

\begin{center}
$Q: State_\alpha \times Order_\alpha \to Query_\alpha$ and $R: Response_\alpha \times State_\alpha \to State_\alpha$.
\end{center}

In the Section \ref{Mode_evaluation} we will introduce the functions for reflexivity. Before that discuss reliability issues.

\subsection{Accuracy and consistency} \label{cdswadkskuew}
In an anaolgue system there are both measured and calculated data. Tracking errors due to measurement, approximation in calculation, time delays, or any other source is vital to the system. Errors are a measure of how well mode $\alpha$ is modelling reality, and they can tell if another mode of operation might do a better job in current circumstances. Changing modes is our primary problem, so 
how do we know that mode $\alpha$ is doing its job of coping with reality? 

Imagine an analogue-digital system with a constant $z$ of the system. Initially we measure $z\in  (0,5)$, and some time later we measure $z\in (4,7)$. In what interval do we now know $z$ to be? It is tempting to say $z\in  (4,5)$, but maybe ...

\smallskip
\noindent \textit{Wear}. The sensor is degrading with time, and that an increase in the midpoint of the interval is part of a gradual drift?

\noindent \textit{Domain}. The day of the second measurement was very hot, and the specification of the sensor says that it should be used at cooler temperatures; that caveat keeps the sensor engineers in the clear, but what about the measurement the system has to work with?

\noindent \textit{Calibration}. The equipment was not recalibrated after a minor shock, and the constant $z$ got shifted.

\smallskip
\noindent Now imagine these complicating factors are not known (e.g., because the system does not have the sensors to report them). Our deduction $z\in  (4,5)$ becomes questionable. We may understand the errors, but do we understand the errors of the errors?\footnote{Compare the idealised situation in computable analysis. Take a computable real number $y$, and suppose we calculate $y\in (0,5)$. Some time later we calculate $y\in (4,7)$, so we deduce that $y\in (4,5)$. As we perform more calculations, the error bound to within which we know $y$ never increases. }

Consider the worse case of a function $f(t)$ of time in an analogue-digital system. We measure $f(0)\in (0,5)$, and the equations of motion given by a theoretical analysis of the system predict $f(1)\in(-2,4)$. We then measure $f(1)\in(3,6)$. We have all the previous complications (1)-(3) above, plus

\smallskip
\noindent \textit{Validity}. Are the equations of motion really valid, or did we go outside their range of validity, or their numerical stability, or are there unexpected \textit{unknown unknowns}?

\smallskip
Surely we want the measured value to be inside the range of the computed value? Measure $g(0)\in(0,2)$, calculate $g(1)\in (-450,900)$ and subsequently measure $g(1)\in (2,4)$. The huge error margin in the calculated value simply illustrates that it is hopelessly inaccurate. The value of all calculated extrapolations tends to garbage as time increases, it is just a matter of how quickly. 

It is important to distinguish between two sorts of \textit{observed} errors, i.e.,\ errors that we know about: If we estimate that the temperature should be in the range $(23. 1,23. 2)$ and it is observed to be in the range 
$(23.4,23. 5)$ then we have an \textit{inconsistency}. However, if we only need to know the temperature to an accuracy of one degree, then the estimate is \textit{accurate}. In the example of $g(t)$ in the last paragraph, we had an example of \textit{consistency} and \textit{inaccuracy}. 

Accuracy matters because being inaccurate may cause a failure of control for the system. Inconsistency matters because it tells us that there is a disagreement between the predicted behaviour and the observed behaviour of the system, and thus that the predicted behaviour should be considered unreliable.

\section{Evaluating a mode implementation}\label{Mode_evaluation}

\subsection{The architecture of mode evaluation}\label{mode_evaluation}
To the four components introduced in subsection \ref{mode_architecture} for gathering information, for each mode $\alpha\in\mathcal{M}$ we now add a fifth for evaluation:

\smallskip
\noindent \textbf{\textsf{monitor}${}_\alpha$}: This measures the suitability of the model and is the only component of mode $\alpha$ which sends information to the system supervisor that decides on mode transitions. 
 
\smallskip
\noindent 
\textbf{Definition.}  The output of the component \textsf{monitor}${}_\alpha$ is specified by a \textit{mode evaluation function} 
\begin{center}
$P_\alpha:\mathcal{M}\times \mathrm{State}_\alpha\times\mathrm{Orders}_\alpha\to [0,1]$. 
\end{center}
\noindent
The number $P_\alpha(\beta,x,o)\in[0,1]$ assesses how well mode $\beta\in\mathcal{M}$ \textit{could} run the system, given the current picture $x\in \mathrm{State}_\alpha$ of the system and the current orders $o\in\mathrm{Orders}_\alpha$. The value of the function is normalised by two given values $0<p_{\mathrm{low}}<p_{\mathrm{high}}<1$ so that

\begin{tabular}{ll}
$P_\alpha(\beta,x,o)=0$ & means mode $\beta$ is \textit{incompatible} with the current state; \\
$P_\alpha(\beta,x,o)<p_{\mathrm{low}}$ & means mode $\beta$ models the system, but \textit{inadequately}; 
\\$P_\alpha(\beta,x,o)>p_{\mathrm{low}}$ & means mode $\beta$ models the system \textit{adequately}; 
\\$P_\alpha(\beta,x,o)>p_{\mathrm{high}}$ & means mode $\beta$ models the system \textit{well}.
\end{tabular}
\newline
\newline
Note that the function $P_\alpha$ is total. The component is responsible for assessing (i) observed problems, such as \textit{consistency} and \textit{accuracy} (Section~\ref{cdswadkskuew}), and (ii) the extent the observed state of the system is consistent with the objectives, including the current orders $o\in\mathrm{Orders}_\alpha$.

\subsection{Simplicial complexes and the geometry of the modes} \label{hadksku}

Next we are concerned with the large scale behaviour of the system and, therefore, the relationships between the modes. 

For some modes $\alpha,\beta,\gamma\in \mathcal{M}$,  the overlap $\alpha\cap\beta\cap\gamma$ might be nonempty. This means, is it possible for the state of the physical system to be simultaneously in (say) all three modes $\alpha,\beta,\gamma$. From such intersection information, we will build a geometric object, called the \textit{nerve} of $\mathcal{M}$ \cite{AlexDim}.\footnote{For the general idea of a nerve of a category rather than a cover by subsets, see \cite{SeaNerve}.} The nerve is a simplicial complex.
\newline
\newline
\noindent 
\textit{Simplicial complexes}.
A \textit{0-simplex} is a point or vertex. A \textit{1-simplex} is a line segment connecting two vertices. A simplicial complex is made of several simplices, so we can see that a simplicial complex consisting of 0-simplices and 1-simplices is just a graph. However, simplicial complexes generalise graphs by allowing higher dimensional constructions. A \textit{2-simplex} is a triangle bounded by three 1-simplices. Figure 5 shows an example of a simplicial complex consisting of one 2-simplex, ten 1-simplices and eight 0-simplices; the 0-simplices are vertices $v$ indexed by $\{\alpha,\beta,\gamma,\delta,\epsilon,\zeta,\theta,\phi\}$.

\unitlength 0.5 mm
\begin{picture}(120,55)(-35,40)
\linethickness{0.3mm}
\multiput(20,80)(0.12,-0.12){250}{\line(1,0){0.12}}
\linethickness{0.3mm}
\put(20,50){\line(1,0){30}}
\linethickness{0.3mm}
\put(20,50){\line(0,1){30}}
\linethickness{0.3mm}
\put(20,80){\line(1,0){40}}
\linethickness{0.3mm}
\multiput(50,50)(0.12,0.36){83}{\line(0,1){0.36}}
\linethickness{0.3mm}
\multiput(60,80)(0.24,-0.12){83}{\line(1,0){0.24}}
\linethickness{0.3mm}
\multiput(80,70)(0.24,0.12){83}{\line(1,0){0.24}}
\linethickness{0.3mm}
\multiput(80,70)(0.12,-0.12){167}{\line(1,0){0.12}}
\linethickness{0.3mm}
\put(100,50){\line(0,1){30}}
\linethickness{0.3mm}
\multiput(100,80)(0.12,-0.12){167}{\line(1,0){0.12}}
\linethickness{0.3mm}
\put(21.88,50){\line(0,1){28.12}}
\linethickness{0.3mm}
\put(23.75,50){\line(0,1){26.25}}
\linethickness{0.3mm}
\put(25.62,50){\line(0,1){24.38}}
\linethickness{0.3mm}
\put(27.5,50){\line(0,1){22.5}}
\linethickness{0.3mm}
\put(29.38,50){\line(0,1){20.62}}
\linethickness{0.3mm}
\put(31.25,50){\line(0,1){18.75}}
\linethickness{0.3mm}
\put(33.12,50){\line(0,1){16.88}}
\linethickness{0.3mm}
\put(35,50){\line(0,1){15}}
\linethickness{0.3mm}
\put(36.88,50){\line(0,1){13.12}}
\linethickness{0.3mm}
\put(38.75,50){\line(0,1){11.25}}
\linethickness{0.3mm}
\put(40.62,50){\line(0,1){9.38}}
\linethickness{0.3mm}
\put(42.5,50){\line(0,1){7.5}}
\linethickness{0.3mm}
\put(44.38,50){\line(0,1){5.62}}
\linethickness{0.3mm}
\put(46.25,50){\line(0,1){3.75}}
\linethickness{0.3mm}
\put(48.12,50){\line(0,1){1.88}}
\put(20,85){\makebox(0,0)[cc]{$v_\alpha$}}

\put(15,45){\makebox(0,0)[cc]{$v_\beta$}}

\put(50,45){\makebox(0,0)[cc]{$v_\gamma$}}

\put(60,85){\makebox(0,0)[cc]{$v_\delta$}}

\put(76,66){\makebox(0,0)[cc]{$v_\epsilon$}}

\put(100,45){\makebox(0,0)[cc]{$v_\zeta$}}

\put(100,85){\makebox(0,0)[cc]{$v_\theta$}}

\put(125,59){\makebox(0,0)[cc]{$v_\phi$}}

%\put(190,45){\makebox(0,0)[cc]{Figure 4}}

\end{picture}

\begin{center}
\textbf{Figure 5:}\quad A simplicial complex
\end{center}

\noindent
The shading indicates that $\alpha\beta\gamma$ is a 2-simplex bounded by the 1-simplices $\alpha\beta$, $\alpha\gamma$ and $\beta\gamma$. Note that there is not a 2-simplex $\alpha\delta\gamma$. The dimensions continue to increase by a \textit{3-simplex} being a tetrahedron (= triangular based pyramid) bounded by four 2-simplex sides, etc. 
Simplicial complexes are used in topology to give a ``computable" presentation of topological spaces \cite{spanier}. 

Coordinates for the simplicial complex are illustrated in Figure 6, which consists of one 2-simplex, four 1-simplices and four 0-simplices $\{\alpha,\beta,\gamma,\delta\}$ (the 2-simplex $\alpha\beta\gamma$ is not shaded for clarity).

\unitlength 0.5 mm
\begin{picture}(130,60)(-30,22)
\linethickness{0.3mm}
\multiput(20,70)(0.12,-0.12){333}{\line(1,0){0.12}}
\linethickness{0.3mm}
\multiput(60,30)(0.12,0.16){250}{\line(0,1){0.16}}
\linethickness{0.3mm}
\put(20,70){\line(1,0){70}}
\linethickness{0.3mm}
\multiput(90,70)(0.12,-0.12){333}{\line(1,0){0.12}}
\put(18,75){\makebox(0,0)[cc]{$v_\alpha$}}

\put(60,25){\makebox(0,0)[cc]{$v_\beta$}}

\put(92,75){\makebox(0,0)[cc]{$v_\gamma$}}

\put(131,25){\makebox(0,0)[cc]{$v_\delta$}}

\put(119,41){\makebox(0,0)[cc]{$\bullet$}}

\put(141,41){\makebox(0,0)[cc]{$\frac13\,v_\gamma\!+\!\frac23\,v_\delta$}}

\put(70,60){\makebox(0,0)[cc]{$\bullet$}}

\put(40,61){\makebox(0,0)[cc]{$\frac14v_\alpha\!+\!\frac14v_\beta\!+\!\frac12v_\gamma$}}

\put(50,40){\makebox(0,0)[cc]{$\bullet$}}

\put(20,41){\makebox(0,0)[cc]{$\frac13v_\alpha\!+\!\frac23v_\beta\!+0\,v_\gamma$}}

%\put(190,25){\makebox(0,0)[cc]{Figure 5}}

\end{picture}

\begin{center}
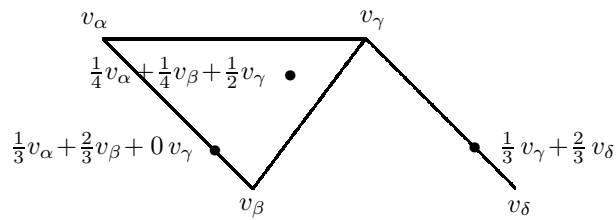
\textbf{Figure 6:}\quad Points in a simplicial complex
\end{center}

\noindent  The 1-simplex $\gamma\delta$ consists of points
\begin{center}
$t_\gamma\,v_\gamma+t_\delta\,v_\delta$ for $t_\gamma,t_\delta\in[0,1]$ and $t_\gamma+t_\delta=1$. 
\end{center} 
We identify the ends of the interval with the vertices, so $1\,v_\gamma+0\,v_\delta$ is identified with $v_\gamma$ and $0\,v_\gamma+1\,v_\delta$ is identified with $v_\delta$.

The  2-simplex $\alpha\beta\gamma$ consists of points
\begin{center}
$t_\alpha\,v_\alpha+t_\beta\,v_\beta+t_\gamma\,v_\gamma$ for $t_\alpha,t_\beta,t_\gamma\in[0,1]$ and $t_\alpha+t_\beta+t_\gamma=1$. 
\end{center}  
We identify the sides of the 2-simplex triangle with the corresponding 1-simplex intervals, for example $\frac13v_\alpha\!+\!\frac23v_\beta\!+0\,v_\gamma$ is identified with $\frac13v_\alpha\!+\!\frac23v_\beta$ in the 1-simplex $\alpha\beta$.\footnote{This sort of coordinate system for a triangle may be compared with the RGB colour system.} 

Note that we do not have $\frac14v_\beta\!+\!\frac14v_\gamma\!+\!\frac12v_\delta$ in the simplicial complex in Figure 5, as 
there is no 2-simplex $\beta\gamma\delta$. 
\newline
\newline
\noindent 
\textit{Construction of nerve}.
The \textit{nerve} $N(\mathcal{M})$ of $\mathcal{M}$ is a simplicial complex defined as follows. 

For every mode $\alpha\in \mathcal{M}$, there is a point or vertex $v_\alpha$. For two different modes $\alpha,\beta\in \mathcal{M}$ with $\alpha\cap\beta$ not empty there is a 1-simplex $\alpha\beta$.

For three different modes $\alpha,\beta,\gamma\in \mathcal{M}$ with $\alpha\cap\beta\cap\gamma$ not empty there is a
2-simplex $\alpha\beta\gamma$, and so on. To summarise this: 
\newline
\newline
\noindent 
\textbf{Definition.} The element $\sum_{\alpha\in \mathcal{M}} t_\alpha\,v_\alpha$ is in the \textit{nerve} $N(\mathcal{M})$ of set $\mathcal{M}$ of modes if, and only if,

(i)  each $t_\alpha\ge 0$ and $\sum_{\alpha\in \mathcal{M}} t_\alpha=1$, and

(ii)  taking the subset of $\alpha\in\mathcal{M}$ for which $t_\alpha> 0$, the intersection of all the modes in the subset is nonempty (i.e.,\ the subset gives a simplex of appropriate size).

\subsection{The mode evaluation and classification functions} \label{mef}
How can there be uncertainty as to which mode the system is in?  Near the boundaries between two subsets it would be common for there to be an overlap. Suppose the program is in mode $\alpha\in \mathcal{M}$.  Let mode $\alpha$ have mode evaluation function
\begin{center}
$P_\alpha:\mathcal{M}\times \mathrm{State}_\alpha\times\mathrm{Orders}_\alpha\to [0,1]$. 
\end{center}
\noindent
Recall from subection \ref{mode_evaluation}, that $P_\alpha(\beta,x,o)$ is a measure of how well $x\in \mathrm{State}_\alpha$ could be described by mode $\beta\in \mathcal{M}$ given the orders $o\in \mathrm{Orders}_\alpha$; and for value $p_{\mathrm{low}}>0$,  if $P_\alpha(\beta,x,o)> p_{\mathrm{low}}$ then $x\in \mathrm{State}_\alpha$ could be reasonably described by mode $\beta$, given orders $o\in \mathrm{Orders}_\alpha$. 
\newline
\newline
\noindent
\textbf{Definition.}The \textit{mode classification function}
\begin{center}
$N_\alpha:\mathrm{State}_\alpha\times\mathrm{Orders}_\alpha\to N(\mathcal{M})$ 
\end{center}
\noindent  is defined as follows: 
\begin{eqnarray} \label{bchlkwkuvcx}
N_\alpha(x,o)\,=\,\frac{\sum_{\beta:P_\alpha(\beta,x,o)> p_{\mathrm{low}}} 
(P_\alpha(\beta,x,o)-p_{\mathrm{low}})\,v_\beta }
{\sum_{\beta:P_\alpha(\beta,x,o)> p_{\mathrm{low}}} (P_\alpha(\beta,x,o)-p_{\mathrm{low}}) }\ .
\end{eqnarray}
The mode classification function \textit{sends the current state $x$ and orders $o$ to a point in the nerve, to give a geometric picture of the situation with respect to the modes}. 
\newline
\newline
Note that we only sum over modes $\beta$ for which $P_\alpha(\beta,x,o)> p_{\mathrm{low}}$, i.e.,\ we only consider those modes which have been certified as being reasonable descriptions of the system, given its observed state. Of course, an objective of the design of the system is that for \textit{any} possible state of the system there is a mode that describes it adequately, and so that there is always a consistent choice of such modes. 
\newline
\newline
\unitlength 0.8 mm


As Figure 7 illustrates, we are considering systems for which there is no unified or global model; the nerve binds the modes together. 

\subsection{Exceptions: known unknowns and unknown unknowns}\label{unknowns}

What could possibly go wrong? Many things could go wrong. Recalling \textit{Principle 0: Robustness}, algorithms ought to respond to \textit{any} eventuality. It is useful to consider Rumsfeld's classification of unknowns. In our situation, and perhaps in system design generally, there are natural working definitions:
\newline
\newline
\noindent 
\textbf{Definition.} A \textit{known unknown} is an event that we knew might happen, we knew the approximate form of the event, and, specifically, that there is software to handle the event. An \textit{unknown unknown} is any event that we did not expect and we do \textit{not} have software to handle. 
\newline
\newline
For an example of a known unknown, in the robotic car race of subsection \ref{carrace}, another car could unexpectedly join the race after it has started; a designer can include a provision for adding cars during the race, which just requires some details to be added if an when the event occurs.  

How do we know when an \textit{unknown unknown} is happening?  The two mathematical singularities that can occur with the mode classification formula (\ref{bchlkwkuvcx}) give a good indication of an observed \textit{unknown unknown}: 
\newline
\newline
\noindent \textit{Partiality.}  The set of modes $\beta$ with $P_\alpha(\beta,x,o)> p_{\mathrm{low}}$ may be empty, i.e.,\ there are no modes which, in the opinion of the designers, model the current state adequately. Since we assumed $p_{\mathrm{low}}\neq 0$, we have some room in which we may choose the best of a lot of bad choices. This is discussed further in Subsection~\ref{ciausciibhaiu}.
\newline
\newline
\noindent \textit{Contradiction.}  For the modes in Figure 5, a classification value of $N_\delta(x,o)$ of $\frac12 v_\gamma+\frac12 v_\delta$ is fine, as there is a line connecting the vertices $v_\gamma$ and $v_\delta$, so the designers knew that those modes could in principle be consistent at a given time. However, the value $\frac12 v_\beta+\frac12 v_\delta$ is invalid, it does not lie in the simplicial complex. This geometric property indicates that the designers never considered the possibility that these modes $v_\beta$ and $v_\delta$ could both be valid descriptions of the same physical state. We have a possible contradiction -- this is discussed further in Subsection~\ref{ciausciibhaiu1}.

\section{Making transitions between modes}\label{Mode_transition}

\subsection{The architecture of mode transition}

There are two components that accomplish a change of mode, one of which was not illustrated earlier in Figure 4:

\smallskip
\noindent\textbf{\textsf{supervisor}}: This component decides on the change of mode from (say) mode $\alpha$ to mode $\beta$. Its input is the value of the mode evaluation function $P_\alpha(\beta,x,o)$ from \textsf{monitor}${}_\alpha$. It assigns the mode $\beta$ to control the system by 

(i) changing state information $\mathrm{State}_\alpha $, and 

(ii) modifying that mode's Orders${}_\alpha$.  

\noindent In the case of observed unknown unknowns it must decide on the system's response (e.g., partiality and consistency -- see subsections~\ref{ciausciibhaiu1} and \ref{ciausciibhaiu}). 

The component is the only part of the program which receives commands from outside agencies; for example, it  is responsible for initialising the whole system on startup; and it can receive an over-ride to move to an emergency shutdown mode. 

\smallskip
\noindent\textbf{\textsf{transfer}${}_\alpha$}: This component moves information between modes -- changing states -- and its specification is the family of \textit{mode transition functions}: 

\begin{center}
$\tau_{\beta\alpha}:\mathrm{State}_\alpha \times\mathrm{Orders}_\alpha\to \mathrm{State}_\beta \times\mathrm{Orders}_\beta$ 
\end{center}

\noindent It is the only part of mode $\alpha$ which needs to know anything about  $\mathrm{State}_\beta$. The mode transition function is partial. When the function $\tau_{\beta\alpha}$ is called by the \textsf{supervisor} and computed, control is transferred to mode $\beta$.

\subsection{The mode transition functions $\tau_{\beta\alpha}$: commenatry} \label{cadshadksku}

If we decide to move from mode $\alpha$ to mode $\beta$, how do we actually implement this? Each mode $\alpha$ carries a picture \textsf{state}${}_\alpha$ of the physical system, so to move from $\alpha$ to $\beta$ we need a mode transition function $\tau_{\beta\alpha}$, so that our picture of reality is redrawn in a form understood by mode $\beta$. However, this function need not be defined on all of State${}_\alpha$; in fact the designers are only likely to have implemented the function $\tau_{\beta\alpha}(x)$ on the condition that $\beta$ is an adequate or good description for $x\in \mathrm{State}_\alpha$. 

The transition functions also depend on the standing orders $\mathrm{Orders}_\alpha$. For example, when controlling temperature, we might simply store the target temperature of a building in the standing orders, e.g.,\ $\mathrm{Orders}_\alpha=\{21^\circ C\}$. In calculating the mode evaluation function $P_\alpha$ we obviously need to take account of the target temperature, but why is it needed in $\tau_{\beta\alpha}$? Simply copying the standing orders, so we would get $\mathrm{Orders}_\beta=\{21^\circ C\}$ may negate the point of changing: the target temperature could be modified by various factors, such as the time of day or the cost of energy, giving e.g.,\ $\mathrm{Orders}_\beta=\{19^\circ C\}$.

Thus, we need the mode transition function to be total when restricted to the form
\begin{eqnarray} \label{bchilsku}
\tau_{\beta\alpha}:\big\{(x,o)\in \mathrm{State}_\alpha\times\mathrm{Orders}_\alpha:P_\alpha(\beta,x,o)>p_{\mathrm{low}}\big\}\to \mathrm{State}_\beta\times \mathrm{Orders}_\beta.
\end{eqnarray}
We need to assume that the designers have done a sufficiently good job of implementing 
$\tau_{\beta\alpha}$ so that the system can carry on from this position in mode $\beta$. If this proves too difficult to do, then it is possible that $\beta$ is not a reasonable description for all $(x,o)\in \mathrm{State}_\alpha\times\mathrm{Orders}_\alpha$ with $P_\alpha(\beta,x,o)>p_{\mathrm{low}}$, and that the function $P_\alpha$ may have to be redefined. 

Note that not all the state variables in $\mathrm{State}_\beta$ need be set by the mode transition function in (\ref{bchilsku}), some may be left undefined. To use the example of the car race in subsection~\ref{carrace}, if we are in mode $\alpha$ we may decide to switch to mode $\beta$ because another car is within a certain distance. However there is no reason why we needed to have much information on that particular car earlier. In particular, we may have no idea of its velocity, or its identity, so these may be left as undefined after the switch in modes. In this case, it is to be hoped that the designers of mode $\beta$ have given a high priority to finding a value for these undefined variables.

One problem with changing modes is what happens when it is done several times in quick succession.
Suppose that we have just used the function in (\ref{bchilsku}) to move from mode $\alpha$ to $\beta$. It may happen that we almost immediately switch to mode $\gamma$. However, the designers could only be expected to have taken account of this possibility if $\gamma$ is a reasonable description of $\tau_{\beta\alpha}(x)\in \mathrm{State}_\beta$. In that case, we get the composition
\begin{eqnarray} \label{bchilsku2}
\tau_{\gamma\beta}\circ\tau_{\beta\alpha}:\big\{y\in \mathrm{State}_\alpha\times\mathrm{Orders}_\alpha :P_\alpha(\beta,y)>p_{\mathrm{low}}\ \mathrm{and}\ P_\beta(\gamma,\tau_{\beta\alpha}(y))>p_{\mathrm{low}}\big\} && \cr
\to \mathrm{State}_\gamma\times\mathrm{Orders}_\gamma.\qquad &&
\end{eqnarray}
However, we might have moved from $\alpha$ to $\gamma$ in one go, using
\begin{eqnarray} \label{bchilsku3}
\tau_{\gamma\alpha}:\big\{y\in \mathrm{State}_\alpha\times\mathrm{Orders}_\alpha:P_\alpha(\gamma,y)>p_{\mathrm{low}}\big\}\to \mathrm{State}_\gamma\times\mathrm{Orders}_\gamma.
\end{eqnarray}
As we could have taken either path from $\alpha$ to $\gamma$,
we should hope that there is a reasonable consistency between $\tau_{\gamma\beta}\circ\tau_{\beta\alpha}$ and 
$\tau_{\gamma\alpha}$ as functions on the subset of $\mathrm{State}_\alpha\times\mathrm{Orders}_\alpha$ where all three inequalities in 
(\ref{bchilsku2}) and (\ref{bchilsku3}) are satisfied; this should be part of the design requirements.

\subsection{Normal running} \label{cdsu}

Changing modes changes the way a system is viewed and may be needed to simplify and frame a computational description. Changing the way a system is viewed comes with a cost: change takes time, it may introduce inaccuracy or inconsistency, and it may lose information. For a robust many mode system, a method is needed by which the \textsf{supervisor} minimises the changes of mode. Periodically, the program checks that it is in a correct mode to describe the system, as follows:

If we are in mode $\alpha$, and $P_\alpha(\alpha,x,o)> p_{\mathrm{high}}$, then $x\in \mathrm{State}_\alpha$ together with the standing orders $o\in\mathrm{Orders}_\alpha$ should be well described by mode $\alpha$. We therefore remain in mode $\alpha$. 

If $P_\alpha(\alpha,x)\le p_{\mathrm{high}}$ we calculate the other values of $P_\alpha(\beta,x,o)$ and use formula (\ref{bchlkwkuvcx}) to find $N_\alpha(x, o)$. We could then select the $\beta\in \mathcal{M}$ which gives 
\[
P_\alpha(\beta,x,o)=\max_\gamma\{P_\alpha(\gamma,x,o)\}\ .
\]
We could modify this in the event of several large values by using the more detailed geometry of the symplectic space 
 by choosing a vertex $v_\beta$ which $N_\alpha(x,o)$ is approaching, comparing the last several values of $N_\alpha(x,o)$. In this manner we pick a mode which the state is entering, and so might be expected to stay in for longer. 

Recalling the \textit{Principle 0: Robustness}, we attempt to implement a \textit{robust} system, with the following working definition:
\newline
\newline
\noindent \textbf{Definition.} The system of modes is \textit{robust} if there are a sufficient number of modes so that we can always choose a mode which models the state well, i.e., with mode evaluation function $>p_{\mathrm{high}}$. 

The system of modes is \textit{adequate} if there are a sufficient number of modes so that we can always choose a mode which models the state adequately, i.e., with mode evaluation function $>p_{\mathrm{low}}$. 
\newline
\newline
These notions reflect the idea that the modularistion of physical behaviour covers all eventualities, normal and exceptional. We can attempt to \textit{verify} that this is the case, by axiomatising physical assumptions about the behaviour of the system and the reliability of our sensors and actuators. 
However, as von Moltke observed, `no plan survives contact with the enemy': our enemy is the physical reality of the system. We need to know when the plan is going wrong, and what to do. In summary, \textit{the plan may fail but the real world continues to move on and decisions have to be taken.}  

We examine the two exceptions which can be thrown by the mode transition function $N_\alpha$, introduced in subsection~\ref{mef}.

\subsection{Partiality exception: Into the unknown} \label{ciausciibhaiu}

In the case of the \textit{partiality exception} in subection~\ref{mef} we have the problem: 
\begin{center}
 $\{\beta\in \mathcal{M}:P_\alpha(\beta,x,o)> p_{\mathrm{low}}  \} = \emptyset$.
 \end{center}

\noindent
In other words, for all $\beta$ we have $P_\alpha(\beta,x,o)\le p_{\mathrm{low}}$, i.e.,\ the software does not think that any mode $\beta$ reasonably describes the picture and orders $(x,o)$. 
 
The assumption, in Section~\ref{hadksku}, that $p_{\mathrm{low}}>0$ means we have some `wriggle room', a gap between where `reasonable description' ends and `complete garbage' begins. We can still try to choose the best of a lot of bad choices -- we may take the largest value of $P_\alpha(\beta,x,o)$ even if it is $\le p_{\mathrm{low}}$. Essentially we may have a mode $\beta$ which the engineers thought was some sort of description of the system, but they were not prepared to certify as a  `reasonable description'.  When using such a mode $\beta$ to control the system, it may be wise to have a \textit{safety strategy} for $\beta$ -- a version of $\mathrm{Orders}_\beta$ written on the understanding that mode $\beta$, while the best available, may still not be a reliable representation of the real world.

But which $\beta$ do we choose? It is possible that the values of $P_\alpha(\beta,x,o)$ may not distinguish well between values in the range $\le p_{\mathrm{low}}$ -- after all, the system was never supposed to be run in that range. When lost but with a last known position mode $\alpha$, we can bias the search for new modes towards ``nearby" modes. The simplicial complex provides an idea of the distance between modes: 
the shortest path in terms of number of edges or 1-simplices between the vertices. 

For example, if for $\alpha,\beta\in \mathcal{M}$ we have a $\gamma$ with $\alpha\cap\gamma\neq\emptyset$ and 
$\gamma\cap\beta\neq\emptyset$ but $\alpha\cap\beta=\emptyset$ then $d(\alpha,\beta)=2$. 
(Alternatively, if the collection of modes $\mathcal{M}$ is part of a multi-scale hierarchy, we might move to a coarser collection of modes: see subsection~\ref{multi-scale}.) 

If the system does not swiftly return to a known mode, and the system has not collapsed, a warning would alert people or another system to intervene.\footnote{The required abstract understanding may well be beyond current artificial intelligence; see \cite{BrownKleerConf} for progress on this front.}

\subsection{Contradiction exception} \label{ciausciibhaiu1}
In the case of the \textit{contradiction exception} in subsection~\ref{mef} we have the problem:

\begin{center}
 $\{\beta\in \mathcal{M}:P_\alpha(\beta,x,o)> p_{\mathrm{low}}  \} \neq \emptyset$ but has no common intersection.
\end{center}

\noindent In the case of three modes $\beta,\gamma,\delta$ with 
$P_\alpha(\beta,x,o)>p_{\mathrm{low}}$, $P_\alpha(\gamma,x,o)>p_{\mathrm{low}}$ and $P_\alpha(\delta,x,o)>p_{\mathrm{low}}$, the following two configurations of intersections of modes in Figure 8 both give problems:

\unitlength 0.5 mm
\begin{picture}(180,80)(0,4)
\linethickness{0.3mm}
\qbezier(30,80)(40.48,77.45)(42.28,72.03)
\qbezier(42.28,72.03)(44.09,66.62)(37.5,57.5)
\qbezier(37.5,57.5)(30.98,48.37)(27.97,44.16)
\qbezier(27.97,44.16)(24.96,39.95)(25,40)
\qbezier(25,40)(25.02,39.93)(23.22,45.34)
\qbezier(23.22,45.34)(21.41,50.76)(17.5,62.5)
\qbezier(17.5,62.5)(13.52,74.26)(16.53,78.47)
\qbezier(16.53,78.47)(19.54,82.68)(30,80)
\linethickness{0.3mm}
\qbezier(30,55)(40.48,52.45)(42.28,46.44)
\qbezier(42.28,46.44)(44.09,40.42)(37.5,30)
\qbezier(37.5,30)(31,19.48)(26.19,20.69)
\qbezier(26.19,20.69)(21.38,21.89)(17.5,35)
\qbezier(17.5,35)(13.52,48.06)(16.53,52.88)
\qbezier(16.53,52.88)(19.54,57.69)(30,55)
\linethickness{0.3mm}
\qbezier(82.5,57.5)(91.7,48.39)(91.09,42.38)
\qbezier(91.09,42.38)(90.49,36.36)(80,32.5)
\qbezier(80,32.5)(69.55,28.51)(65.34,32.72)
\qbezier(65.34,32.72)(61.13,36.93)(62.5,50)
\qbezier(62.5,50)(63.75,63.1)(68.56,64.91)
\qbezier(68.56,64.91)(73.38,66.71)(82.5,57.5)
\linethickness{0.3mm}
\qbezier(160,75)(170.5,72.43)(170.5,68.22)
\qbezier(170.5,68.22)(170.5,64.01)(160,57.5)
\qbezier(160,57.5)(149.6,51.02)(141.78,45)
\qbezier(141.78,45)(133.96,38.98)(127.5,32.5)
\qbezier(127.5,32.5)(120.98,25.93)(117.97,26.53)
\qbezier(117.97,26.53)(114.96,27.13)(115,35)
\qbezier(115,35)(114.99,42.8)(115.59,48.22)
\qbezier(115.59,48.22)(116.2,53.63)(117.5,57.5)
\qbezier(117.5,57.5)(118.77,61.39)(121.78,65)
\qbezier(121.78,65)(124.79,68.61)(130,72.5)
\qbezier(130,72.5)(135.16,76.43)(142.38,77.03)
\qbezier(142.38,77.03)(149.59,77.63)(160,75)
\linethickness{0.3mm}
\qbezier(147.5,35)(151.51,29.77)(146.09,28.56)
\qbezier(146.09,28.56)(140.68,27.36)(125,30)
\qbezier(125,30)(109.32,32.59)(103.91,35.59)
\qbezier(103.91,35.59)(98.49,38.6)(102.5,42.5)
\qbezier(102.5,42.5)(106.37,46.42)(111.78,47.62)
\qbezier(111.78,47.62)(117.2,48.83)(125,47.5)
\qbezier(125,47.5)(132.8,46.23)(138.22,43.22)
\qbezier(138.22,43.22)(143.63,40.21)(147.5,35)
\linethickness{0.3mm}
\qbezier(150,50)(144.72,31.73)(147.12,23.91)
\qbezier(147.12,23.91)(149.53,16.09)(160,17.5)
\qbezier(160,17.5)(170.47,18.7)(172.88,27.12)
\qbezier(172.88,27.12)(175.28,35.55)(170,52.5)
\qbezier(170,52.5)(164.81,69.57)(160,68.97)
\qbezier(160,68.97)(155.19,68.37)(150,50)
\put(25,70){\makebox(0,0)[cc]{$\beta$}}

\put(130,60){\makebox(0,0)[cc]{$\beta$}}

\put(25,30){\makebox(0,0)[cc]{$\gamma$}}

\put(138,33){\makebox(0,0)[cc]{$\gamma$}}

\put(70,50){\makebox(0,0)[cc]{$\delta$}}

\put(160,35){\makebox(0,0)[cc]{$\delta$}}

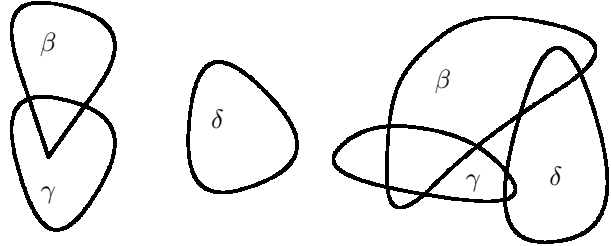
\put(110,10){\makebox(0,0)[cc]{\textbf{   Figure 8:} Intersections of modes  }}

\end{picture}

%\noindent \textbf{Figure 8:} Intersections of modes

\smallskip
We have statements that $(x,o)\in \mathrm{State}_\alpha\times \mathrm{Orders}_\alpha$ could be reasonably described by a collection of modes which has no common intersection. But surely this is a real contradiction, as $(x,o)$ would be in the intersection? The answer is no, and the reason is lack of information or errors. Remember that the point of modes is that \textsf{state}${}_\alpha$ is \textit{not} the state space of the whole system, but rather a picture created by the software to describe one aspect, mode  $\alpha$,
 of the system. 
Let us digress with a new scenario to illustrate this:

\smallskip
\textit{Example: Aircraft}. A military aircraft spots a new contact on its radar screen. There is, as yet, insufficient information to say if it is a hostile military plane or a civilian plane. Both possibilities \textit{could} reasonably describe the situation as either might well be true or false; but, in particular, there is no possibility that describes the situation well. If we just have the two disjoint modes $\alpha=\mathrm{hostile}$ and $\beta=\mathrm{civilian}$ to choose from,
we have the case where there is more than one possible mode assignment consistent with the data (not necessarily a problem), but the assignments are, in reality, mutually exclusive (a problem). The reason is lack of data. We know that the designers did not consider this case, as if so they would have drawn a line joining $\alpha$ and $\beta$ in the simplicial complex, and no exception would have occurred. Of course, it might be said that the designers \textit{should} have considered this case, and that would be a valid point to make at a committee of enquiry into any disaster resulting from the situation, but is not much help at the time. 

\smallskip
So we know there is a problem, but what do we do? We can flag an error, and hope that a human puts the matter right before something unfortunate happens -- if the system can simply be shut down in a safe mode. Extra information could resolve the problem, and the exception should trigger several queries for more information. But what do we do \textit{now}? Here are four options that might work in the airplane scenario:

\smallskip
\noindent \textit{Hawk}. Knowing that there is a significant chance it might be wrong, the software takes the most likely mode, i.e., the largest $P_\alpha$, and, e.g., as a result fires missiles to shoot the target down. For example, in the absence of other information, the matter was decided because the unknown plane was not on a commercial flight path. 

\smallskip
\noindent \textit{Dove}. Again knowing that there is a significant chance it might be wrong, the software takes the most likely mode. But although the plane may assume that the other aircraft is hostile $\alpha$, it is operating under safety strategy  where standing orders are used when the mode is active but considered to be unreliable. In this case an obvious thing to write in $\mathrm{Orders}_\alpha$ would be `do not fire missiles'. 

\smallskip
\noindent \textit{Fail-safe}. Written into the code for $P_\alpha$ is the guarantee: 
\begin{center}
$P_\alpha(\textrm{civilian},x,o)> p_{\mathrm{low}}$ $\Longrightarrow$ $P_\alpha(\textrm{hostile},x,o)< P_\alpha(\textrm{civilian},x,o)$. 
\end{center}

\noindent This means that a taking the largest choice defaults to civilian rather than hostile, in the case where the situation is reasonably described by the unknown plane being civilian. Thus, if we fail to identify the plane successfully, the default is fail-safe (well, as far as the possible civilians are concerned). Of course there is a slight inconsistency here -- if the engineers rigged this fail-safe, why did they not simply include code for handling an intersection of hostile and civilian? 

\smallskip
\noindent \textit{Fail-safe: Consensus}. During the design process the two design teams for the modes are asked: `\textit{Suppose that we have an inconsistent mode allocation, but the evidence indicates that your mode may be applicable. What orders should be issued as a result, given that your mode may equally well not be applicable?}'. The civilian mode designers might reply
`do not fire missiles', and the hostile mode designers might reply `take evasive action to avoid attack'. Now whichever mode takes control of the system is given \textit{consensus} orders containing both the orders above. This is effectively a way of making up some sort of fail-safe strategy as the design develops.

\section{Systems modelled by manifolds and geodesics} \label{cadsbqil}

Many physical systems are modelled by dynamical systems based on differential manifolds \cite{RiemGeom,Arnold_ODE,Arnold_CMech,SmaleODE}.  A manifold is a topological space whose neighbourhoods have local coordinate systems based on $\mathbb{R}^n$. These local coordinate systems are called \textit{charts}. In fact, a manifold is \textit{determined} by its charts. Each chart is a coordinate system optimised for the description of a dynamical behaviour in its neighbourhood. The behaviour of the dynamical system in time is a path through the manifold passing through the charts. Thus, the manifold is a mathematical construction that combines distinct local representations into a global representation of states and behaviours.  

If a complex system can be modelled by a manifold then we are in a special case where there are many tools that can be applied to analyse the system. We will show how 

(i) modes can be based upon the local representations by charts, and 

(ii) the modes change as the behaviour of the dynamical system changes in time. 

Such dynamical systems are common. They are, however, an important example as our theory of multi-modal analogue digital systems is specifically designed so as \textit{not} require there to be a global state space for the physical system.

\subsection{Smooth manifolds}\label{manifolds}

A topological space $X$ is an $n$-dimensional smooth manifold if it possess a family of charts. Let $\mathcal{M}$ be some index set. According to the usual definition, a \textit{chart} consists of 
\begin{center}
open subsets $U_\alpha \subset  X$ and homeomorphisms $\phi_\alpha:U_\alpha\to V_\alpha$, 
\end{center}
\noindent where each $V_\alpha$ is an open subset of $\mathbb{R}^n$ and $\alpha \in \mathcal{M}$. 

Furthermore, for every $\alpha,\beta\in \mathcal{M}$ with $U_\alpha\cap U_\beta$ not empty, the \textit{transition function} 
\begin{eqnarray}
\tau_{\beta\alpha}=\phi_\beta\,\phi_\alpha^{-1}: (\phi_\alpha(U_\alpha\cap U_\beta)\subset V_\alpha)\to
(\phi_\beta(U_\alpha\cap U_\beta)\subset V_\beta)
\end{eqnarray}
is a smooth function (i.e.,\ differentiable arbitrarily many times). Further if $U_\alpha\cap U_\beta \cap U_\gamma$ is not empty we have the equality
\begin{eqnarray}
\tau_{\gamma\alpha}=\tau_{\gamma\beta}\,\tau_{\beta\alpha}: (\phi_\alpha(U_\alpha\cap U_\beta\cap U_\gamma)\subset V_\alpha)\to
(\phi_\gamma(U_\alpha\cap U_\beta\cap U_\gamma)\subset V_\gamma)\ .
\end{eqnarray}

To compute with this we need a function to tell us which chart we are in. This is given by a \textit{partition of unity} \cite{RudRealComp} p.\ 40, which is a collection of functions: 
\begin{center}
$\theta_\alpha:X\to [0,1]$ with $\sum_\alpha \theta_\alpha(x)=1$ for all $x\in X$ and $\theta_\alpha(x)=0$ if $x\notin U_\alpha$. 
\end{center}
\noindent (Thus we have a `part'  of  `unity' 1 for each open set.) We relate these functions to the charts $V_\alpha\subset \mathbb{R}^n$ by defining
\begin{eqnarray}
P_\alpha(\beta)=\theta_\beta\,\phi_\alpha^{-1}:V_\alpha\to [0,\infty)\ .
\end{eqnarray}
Figure 9 illustrates the nerve for a covering of a manifold with charts labelled by $\{\alpha,\beta,\gamma,\delta\}$ (the triangle is not shaded for clarity). Keeping this order, the function 
 $(P_\alpha(\alpha), P_\alpha(\beta),P_\alpha(\gamma),P_\alpha(\delta))$ gives a function into the nerve, for illustration mapping the displayed point on the manifold (in $\alpha$ coordinates) to the point on the nerve.

For purposes of computation it is convenient to extend this $P_\alpha(\beta)=\theta_\beta\,\phi_\alpha^{-1}:V_\alpha\to [0,\infty)\ $  to a function on domain $\mathbb{R}^n$ by taking the function to be zero outside $V_\alpha$.

 \unitlength 0.5 mm


 \medskip

If there are originally $m$ open sets, then if we choose $p_{\mathrm{low}}=\frac1{m+1}$, for any $\alpha$ and $y\in V_\alpha$ we are guaranteed to have at least one $\beta$ so that $P_\alpha(\beta,y)>p_{\mathrm{low}}$.\footnote{A better bound for $p_{\mathrm{low}}$ in terms of the dimension is possible from the 
 \u{C}ech-Lebesgue covering dimension \cite{HurWal}.} Recalling the definition in subsection \ref{cdsu}, we have:
\newline
\newline
\noindent \textbf{Lemma.} \textit{The construction of an analogue-digital system with finitely many modes based upon the finitely many charts of a differential manifold as specified above is adequate.}
\newline
\newline
The theory of manifolds is extensive and contains many intriguing results about inequivalent families of charts and the embedding of $n$-dimensional manifolds in sufficiently higher dimension spaces $\mathbb{R}^N$. The question as to whether and how computation on manifolds should be split into coordinate charts -- rather than, e.g.,\ regarding it as s subset of a higher dimensional Euclidean space -- is answered in the extensive and closely related literature on the choice of grids for global climate models, e.g.,\ \cite{AdCaHiMa}.

\subsection{A dynamical system}\label{dynamical_system}
Consider how a practical problem involves the idea of a manifold. The manifold represents numerically and geometrically the space of all possible states of a system. The behaviour of the system in time is represented -- geometrically and numerically -- by a vector field and paths in the manifold. There is a differential equation governing the time evolution of the coordinates of states, which are points in the path traversing the manifold. Periodically, we have to \textit{change} coordinate charts. 

For example, consider the idea of a \textit{geodesic} or minimum distance path on a curved surface; it has been likened to the path of an ant crawling on the surface of an apple \cite{MSWgrav}. In Einstein's general theory of relativity \cite {EinGR} the paths of the planets are given by geodesics in a space-time manifold, and the resulting prediction of the perihelion precession of Mercury was central in convincing people about that theory. (See  \cite{MSWgrav,RindEssRel} for the mathematical details.)

The definition of a manifold and geodesic motion fits our idea of computation of many modes as follows. To begin, we assume that we have a register machine with real number registers, and operations that enable it 

(i) to calculate precisely all those inconvenient transition functions between charts; and even 

(ii) to solve the differential equations for the geodesics.   

\noindent The modes are exactly the set of charts indexed by $\mathcal{M}$. The picture of the system corresponding to each mode $\alpha \in \mathcal{M}$ is $\mathrm{State}_\alpha=\mathbb{R}^n\times \mathbb{R}^n$. We consider $(y(t),\dot y(t))\in 
\mathbb{R}^n\times \mathbb{R}^n$, where $y(t)$ is the state (e.g., position of our ant, airplane or planet) at time $t$ in the coordinate chart $V_\alpha$, and $\dot y(t)$ is its rate of change or velocity. No attempt to place any restriction forcing $y(t)\in V_\alpha$ is needed: it is the job of the functions $P_\alpha(\beta,(q,p))=\theta_\beta\,\phi_\alpha^{-1}(q)$ to ensure that charts are changed as appropriate. (This type of example does not include $\mathrm{Orders}_\alpha$ as the program only has one purpose, with no variability.)

Next, we consider how these functions are computed. In computable analysis, computable reals and computable functions of computable reals are computed by programs generating rational numbers with error bars \cite{WeiCompAn}. To find the position at a given time to a given error, we have to find how accurately we need to know the initial point, and then time step into the future, keeping track of errors and (rather messily) changing charts where we have to. There may be singularities, places beyond which we cannot extend the geodesics, but bearing that in mind, this is a perfectly consistent computable mathematical theory. However, we no longer have $\mathrm{State}_\alpha=\mathbb{R}^n\times \mathbb{R}^n$, we need a more complicated data type incorporating error bounds for $(y(t),\dot y(t))$. (For analogue-digital systems see \ \cite{HicWitt}.) 

What complications can arise between our mathematical model and the observations? 
\newline
\newline
\noindent \textit{Overdue computation}. The real physical system waits for no computer, it evolves regardless of whether the computer has finished a particular calculation. In computable analysis, if the velocity becomes too large we can reduce the time-step in the program to compensate. In real physical time we may not have this option, and the velocity may become so large that we even lose track of which chart or mode the system is in.
 
\smallskip
\noindent \textit{Chaotic computation}. Many physical systems are chaotic, effectively meaning that even if we had an arbitrarily large amount of time to compute and an exact physical theory, we still could not measure the initial conditions exactly enough to extrapolate beyond a certain time limit. Of course, this could be a \textit{known unknown}, and we can force the system to compensate by regularly observing the most chaotic variables, and using these to correct the calculated values. In the case of the solar system, this would be the positions of the planets around their orbits, as opposed to the much more slowly changing orbital parameters (for the solar system see \cite{laskChaos}).

\section{A racing track} \label{chicane}

\subsection{A single car}

Consider a toy car race around the track shown in Figure 10:
\newline
\newline
\unitlength 0.6 mm


%\noindent \textbf{Figure 9:} A racing track

\smallskip
\noindent
\textbf{Mode $\alpha$: Free running}: 
For a single car, the slots on the track keeps the car moving in a given path, and all we have to worry about is getting the fastest time round the track, taking account of the problem that the maximum speed of the car is too fast for the semicircular parts of the track, so the car has to slow down there. For car $i = 1, 2$ this mode $\alpha^i$ has $\mathrm{State}^i_{\alpha}=(q_i,v_i)$, where $q_i$ is the position of car $i$ and $v_i$ is the velocity of car $i$. 

\smallskip
\noindent
\textbf{Mode $\beta$: Near chicane}: 
When two cars $1$ and $2$ are on the track, they can collide where the track narrows at the chicane, at position $q_c$. To stop this happening, a new mode $\beta^i$ is introduced for car $i$ for collision avoidance. We need data on both cars, so 
$\mathrm{State}^i_{\beta}=(q_1,v_1,q_2,v_2)$. 

\smallskip
\noindent
\textbf{Mode evaluation}: 
For simplicity we assume that mode $\beta^i$ is activated any time that car $i$ is near the chicane. We have \textit{mode evaluation functions}
\begin{eqnarray*}
&& P^i_{\alpha}(\alpha^i,(q_i,v_i))=f(|q_i-q_c|)\ ,\cr
&& P^i_{\alpha}(\beta^i,(q_i,v_i))=1-f(|q_i-q_c|)\ ,\cr
&& P^i_{\beta}(\alpha^i,(q_1,v_1,q_2,v_2))=f(|q_i-q_c|)\ ,\cr
&& P^i_{\beta}(\beta^i,(q_1,v_1,q_2,v_2))=1-f(|q_i-q_c|)\ ,
\end{eqnarray*}
where $|q_i-q_c|$ is the distance from car $i$ to the chicane, $d$ is chosen according to the stopping distance of the car, and $f$ is given by the graph
in Figure 11:

\unitlength 0.5 mm


%\noindent \textbf{Figure 10:} The graph of $f$

\smallskip

\noindent
\textbf{The nerve}: This consists of the $0$-simplices $\alpha$ and $\beta$, and the 1-simplex $(\alpha,\beta)$.

\unitlength 0.5 mm
%

%\noindent \textbf{Figure 11:} Single car nerve for the car race

\smallskip

\noindent
\textbf{Transition functions}: We suppose that Orders is trivial for simplicity. Then 

\smallskip $\tau^1_{\alpha,\beta}:
\mathrm{State}^1_{\beta} \to \mathrm{State}^1_{\alpha}$ is given by $\tau^1_{\alpha,\beta}(q_1,v_1,q_2,v_2)=(q_1,v_1)$ and 

\smallskip 
$\tau^1_{\beta,\alpha}:
\mathrm{State}^1_{\alpha} \to \mathrm{State}^1_{\beta}$ is given by $\tau^1_{\beta,\alpha}(q_1,v_1)=(q_1,v_1,?,?)$. \newline
 
\noindent Here $?$ denotes an undefined value, and one which it is a high priority to find out about.

\subsection{Several cars}
Implementing the safety measures in the car race depends on knowing the positions of the other cars. This is likely to involve other processors, as 
competing teams would likely be unhappy at the thought of a single process controlling both team's cars. In other words, we have a network. Just how much cooperation between nodes of the network do we need to ensure that our safety measures work? 

\smallskip\noindent\textbf{Autonomous}: A team fitted their car with a camera to observe nearby cars. They can now implement mode $\beta$ entirely independently of what any other team does.

\smallskip\noindent\textbf{Communal}: The race organisers have made a condition of entry to the competition, that all cars must make their position public if they are near the chicane, in order to avoid accidents. If car 1 enters mode $\beta^1$ it queries car 2, and if car 2 is also in mode $\beta^2$ it is obliged to respond. 

\smallskip\noindent\textbf{Priority}: A particular team has incorporated into mode $\beta$ software to maximise their team points in the competition, in the case that two of its cars are racing each other. For example, one car could let the other always go first into the chicane. 
If possible, the team might actually combine the processes $\beta^1$ and $\beta^2$ into a single process $\beta^{12}$ running on a single computer.

\section{A solar system} \label{solar}

\subsection{Real time modelling of a solar system}
The Sun, Moon and Earth form a three body gravitational system, and there is no exact general solution known to the three body gravitational problem. The point is that we do not have a general three body problem with the Sun, Moon and Earth: To a very good approximation the Moon orbits the Earth, and the Earth-Moon system orbits the Sun, and our computational problem is vastly simplified.

More generally, take an example of a solar system with three planets $\{1,2,3\}$ orbiting a star. As long as the planets are a reasonable distance apart, the dominant feature in their orbits is the star's gravity, with minor influences from the other planets. This is modelled by processes or modes $\{\alpha^1,\alpha^2,\alpha^3\}$, which can easily be run on different processors or nodes, as occasional communication of the other planets orbits is sufficient to keep the planet $i$'s orbit updated. 

\smallskip
\noindent \textbf{Modes}: Let us concentrate on planet 1. If one of the other planets, say 2, comes within a certain distance of 1, then we require much more detailed knowledge of the position and velocity of planet 2 to accurately model the path of planet 1. We will call this mode $\beta^1_2$ (reserving the superscript $i$ for a process run on node $i$). 
Likewise there is a mode $\beta^1_3$ used when planet 3 is close to planet 1, and a mode $\gamma^1_{23}$ when both other planets are close to 1. 
Thus node 1 has four modes of operation, $\mathcal{M}^1=\{\alpha^1,\beta^1_2,\beta^1_3,\gamma^1_{23}\}$. 

\smallskip
\noindent
\textbf{Mode evaluation}: 
To choose between these modes, we use indicator functions $\{w^1_2,w^1_3\}$. The definition of these is quite simple, the value $w^1_i\in[0,1]$ and $w^1_i\cong 0$ when planet $i$ is `far' from planet 1, and $w^1_i\cong 1$ when planet $i$ is `close' to planet 1. Thus $w^1_i$ may be thought of as a fuzzy logic truth value of the statement \textit{planet $i$ is close to planet 1}. 

\unitlength 0.4 mm
\begin{picture}(140,95)(0,-10)
\linethickness{0.3mm}
\put(42.5,44.75){\line(0,1){0.49}}
\multiput(42.46,44.26)(0.03,0.49){1}{\line(0,1){0.49}}
\multiput(42.4,43.78)(0.06,0.49){1}{\line(0,1){0.49}}
\multiput(42.3,43.3)(0.1,0.48){1}{\line(0,1){0.48}}
\multiput(42.18,42.82)(0.13,0.47){1}{\line(0,1){0.47}}
\multiput(42.02,42.36)(0.16,0.46){1}{\line(0,1){0.46}}
\multiput(41.83,41.9)(0.09,0.23){2}{\line(0,1){0.23}}
\multiput(41.61,41.46)(0.11,0.22){2}{\line(0,1){0.22}}
\multiput(41.37,41.04)(0.12,0.21){2}{\line(0,1){0.21}}
\multiput(41.1,40.63)(0.14,0.2){2}{\line(0,1){0.2}}
\multiput(40.8,40.24)(0.15,0.19){2}{\line(0,1){0.19}}
\multiput(40.47,39.87)(0.11,0.12){3}{\line(0,1){0.12}}
\multiput(40.13,39.53)(0.12,0.12){3}{\line(1,0){0.12}}
\multiput(39.76,39.2)(0.12,0.11){3}{\line(1,0){0.12}}
\multiput(39.37,38.9)(0.19,0.15){2}{\line(1,0){0.19}}
\multiput(38.96,38.63)(0.2,0.14){2}{\line(1,0){0.2}}
\multiput(38.54,38.39)(0.21,0.12){2}{\line(1,0){0.21}}
\multiput(38.1,38.17)(0.22,0.11){2}{\line(1,0){0.22}}
\multiput(37.64,37.98)(0.23,0.09){2}{\line(1,0){0.23}}
\multiput(37.18,37.82)(0.46,0.16){1}{\line(1,0){0.46}}
\multiput(36.7,37.7)(0.47,0.13){1}{\line(1,0){0.47}}
\multiput(36.22,37.6)(0.48,0.1){1}{\line(1,0){0.48}}
\multiput(35.74,37.54)(0.49,0.06){1}{\line(1,0){0.49}}
\multiput(35.25,37.5)(0.49,0.03){1}{\line(1,0){0.49}}
\put(34.75,37.5){\line(1,0){0.49}}
\multiput(34.26,37.54)(0.49,-0.03){1}{\line(1,0){0.49}}
\multiput(33.78,37.6)(0.49,-0.06){1}{\line(1,0){0.49}}
\multiput(33.3,37.7)(0.48,-0.1){1}{\line(1,0){0.48}}
\multiput(32.82,37.82)(0.47,-0.13){1}{\line(1,0){0.47}}
\multiput(32.36,37.98)(0.46,-0.16){1}{\line(1,0){0.46}}
\multiput(31.9,38.17)(0.23,-0.09){2}{\line(1,0){0.23}}
\multiput(31.46,38.39)(0.22,-0.11){2}{\line(1,0){0.22}}
\multiput(31.04,38.63)(0.21,-0.12){2}{\line(1,0){0.21}}
\multiput(30.63,38.9)(0.2,-0.14){2}{\line(1,0){0.2}}
\multiput(30.24,39.2)(0.19,-0.15){2}{\line(1,0){0.19}}
\multiput(29.87,39.53)(0.12,-0.11){3}{\line(1,0){0.12}}
\multiput(29.53,39.87)(0.12,-0.12){3}{\line(0,-1){0.12}}
\multiput(29.2,40.24)(0.11,-0.12){3}{\line(0,-1){0.12}}
\multiput(28.9,40.63)(0.15,-0.19){2}{\line(0,-1){0.19}}
\multiput(28.63,41.04)(0.14,-0.2){2}{\line(0,-1){0.2}}
\multiput(28.39,41.46)(0.12,-0.21){2}{\line(0,-1){0.21}}
\multiput(28.17,41.9)(0.11,-0.22){2}{\line(0,-1){0.22}}
\multiput(27.98,42.36)(0.09,-0.23){2}{\line(0,-1){0.23}}
\multiput(27.82,42.82)(0.16,-0.46){1}{\line(0,-1){0.46}}
\multiput(27.7,43.3)(0.13,-0.47){1}{\line(0,-1){0.47}}
\multiput(27.6,43.78)(0.1,-0.48){1}{\line(0,-1){0.48}}
\multiput(27.54,44.26)(0.06,-0.49){1}{\line(0,-1){0.49}}
\multiput(27.5,44.75)(0.03,-0.49){1}{\line(0,-1){0.49}}
\put(27.5,44.75){\line(0,1){0.49}}
\multiput(27.5,45.25)(0.03,0.49){1}{\line(0,1){0.49}}
\multiput(27.54,45.74)(0.06,0.49){1}{\line(0,1){0.49}}
\multiput(27.6,46.22)(0.1,0.48){1}{\line(0,1){0.48}}
\multiput(27.7,46.7)(0.13,0.47){1}{\line(0,1){0.47}}
\multiput(27.82,47.18)(0.16,0.46){1}{\line(0,1){0.46}}
\multiput(27.98,47.64)(0.09,0.23){2}{\line(0,1){0.23}}
\multiput(28.17,48.1)(0.11,0.22){2}{\line(0,1){0.22}}
\multiput(28.39,48.54)(0.12,0.21){2}{\line(0,1){0.21}}
\multiput(28.63,48.96)(0.14,0.2){2}{\line(0,1){0.2}}
\multiput(28.9,49.37)(0.15,0.19){2}{\line(0,1){0.19}}
\multiput(29.2,49.76)(0.11,0.12){3}{\line(0,1){0.12}}
\multiput(29.53,50.13)(0.12,0.12){3}{\line(0,1){0.12}}
\multiput(29.87,50.47)(0.12,0.11){3}{\line(1,0){0.12}}
\multiput(30.24,50.8)(0.19,0.15){2}{\line(1,0){0.19}}
\multiput(30.63,51.1)(0.2,0.14){2}{\line(1,0){0.2}}
\multiput(31.04,51.37)(0.21,0.12){2}{\line(1,0){0.21}}
\multiput(31.46,51.61)(0.22,0.11){2}{\line(1,0){0.22}}
\multiput(31.9,51.83)(0.23,0.09){2}{\line(1,0){0.23}}
\multiput(32.36,52.02)(0.46,0.16){1}{\line(1,0){0.46}}
\multiput(32.82,52.18)(0.47,0.13){1}{\line(1,0){0.47}}
\multiput(33.3,52.3)(0.48,0.1){1}{\line(1,0){0.48}}
\multiput(33.78,52.4)(0.49,0.06){1}{\line(1,0){0.49}}
\multiput(34.26,52.46)(0.49,0.03){1}{\line(1,0){0.49}}
\put(34.75,52.5){\line(1,0){0.49}}
\multiput(35.25,52.5)(0.49,-0.03){1}{\line(1,0){0.49}}
\multiput(35.74,52.46)(0.49,-0.06){1}{\line(1,0){0.49}}
\multiput(36.22,52.4)(0.48,-0.1){1}{\line(1,0){0.48}}
\multiput(36.7,52.3)(0.47,-0.13){1}{\line(1,0){0.47}}
\multiput(37.18,52.18)(0.46,-0.16){1}{\line(1,0){0.46}}
\multiput(37.64,52.02)(0.23,-0.09){2}{\line(1,0){0.23}}
\multiput(38.1,51.83)(0.22,-0.11){2}{\line(1,0){0.22}}
\multiput(38.54,51.61)(0.21,-0.12){2}{\line(1,0){0.21}}
\multiput(38.96,51.37)(0.2,-0.14){2}{\line(1,0){0.2}}
\multiput(39.37,51.1)(0.19,-0.15){2}{\line(1,0){0.19}}
\multiput(39.76,50.8)(0.12,-0.11){3}{\line(1,0){0.12}}
\multiput(40.13,50.47)(0.12,-0.12){3}{\line(0,-1){0.12}}
\multiput(40.47,50.13)(0.11,-0.12){3}{\line(0,-1){0.12}}
\multiput(40.8,49.76)(0.15,-0.19){2}{\line(0,-1){0.19}}
\multiput(41.1,49.37)(0.14,-0.2){2}{\line(0,-1){0.2}}
\multiput(41.37,48.96)(0.12,-0.21){2}{\line(0,-1){0.21}}
\multiput(41.61,48.54)(0.11,-0.22){2}{\line(0,-1){0.22}}
\multiput(41.83,48.1)(0.09,-0.23){2}{\line(0,-1){0.23}}
\multiput(42.02,47.64)(0.16,-0.46){1}{\line(0,-1){0.46}}
\multiput(42.18,47.18)(0.13,-0.47){1}{\line(0,-1){0.47}}
\multiput(42.3,46.7)(0.1,-0.48){1}{\line(0,-1){0.48}}
\multiput(42.4,46.22)(0.06,-0.49){1}{\line(0,-1){0.49}}
\multiput(42.46,45.74)(0.03,-0.49){1}{\line(0,-1){0.49}}

\put(25,70){\makebox(0,0)[cc]{$\bullet_1$}}
\put(15,60){\makebox(0,0)[cc]{$\bullet_2$}}
\put(95,20){\makebox(0,0)[cc]{$\bullet_3$}}

\put(56,45){\makebox(0,0)[cc]{Star}}

\put(90,0){
\linethickness{0.3mm}
\put(90,70){\line(1,0){50}}
\put(90,20){\line(0,1){50}}
\put(140,20){\line(0,1){50}}
\put(90,20){\line(1,0){50}}
\put(115,12){\makebox(0,0)[cc]{$w^1_2$}}

\put(81,45){\makebox(0,0)[cc]{$w^1_3$}}

\put(95,25){\makebox(0,0)[cc]{0}}

\put(135,25){\makebox(0,0)[cc]{1}}

\put(95,65){\makebox(0,0)[cc]{1}}

\put(70,18){\makebox(0,0)[cc]{$(1,0,0,0)$}}
\put(70,72){\makebox(0,0)[cc]{$(0,0,1,0)$}}
\put(161,18){\makebox(0,0)[cc]{$(0,1,0,0)$}}
\put(161,72){\makebox(0,0)[cc]{$(0,0,0,1)$}}
\put(128,29){\makebox(0,0)[cc]{$\bullet$}}
}

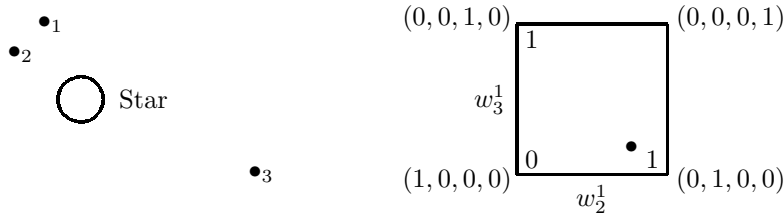
\put(125,0){\makebox(0,-2)[cc]{\textbf{Figure 13:} Configurations of planets}}

\end{picture}

%\noindent \textbf{Figure 12:} Configurations of planets

\smallskip

\noindent
The indicator functions $(w^1_2,w^1_3)$ are converted to mode evaluation functions by
\begin{eqnarray} \label{welldescplanet}
(w^1_2,w^1_3) \longmapsto \big((1-w^1_2)(1-w^1_3),\,w^1_2\,(1-w^1_3),\,(1-w^1_2)\,w^1_3,\, w^1_2\,w^1_3
\big)\ ,
\end{eqnarray}
where we use the order $\{\alpha^1,\beta^1_2,\beta^1_3,\gamma^1_{23}\}$.
In Figure 13 we have a configuration of planets with, say as indicated by the dot on the
square of  $(w^1_2,w^1_3)$ values,    $(w^1_2,w^1_3)=(0.8,0.2)$, giving mode evaluation functions
$(0.16, 0.64, 0.04,0.16)$. Thus the most suitable mode for node 1 in Figure 13  is $\beta^1_2$. Figure 13 also shows the value of the mode evaluation functions for the corners of the square of  $(w^1_2,w^1_3)$ values. 

\smallskip
\noindent
\textbf{The nerve}: This is the solid tetrahedron with vertices $\{\alpha^1,\beta^1_2,\beta^1_3,\gamma^1_{23}\}$. It is a 3-simplex because, as shown in the example, we expect that all four mode evaluation functions can be nonzero simultaneously. 

\smallskip
\noindent
\textbf{Transition functions}: We have not specified the data structures associated to the modes, however if the modes $\beta$ used Einstein's general relativity and mode $\alpha$ used Newton's theory, it might be imagined that quite a bit of data conversion would have to be done. 

\subsection{Cooperation between processes tracking planets}
 It was found that communication problems between different machines meant that running two separate processes  for planets which were close together was too inaccurate (evidently this star system is more interesting and dangerous than ours). It was decided to instigate a cooperative process
$\mathcal{M}^{12}\subset \mathcal{M}^{1}\times \mathcal{M}^{2}$, consisting of just one mode
$\beta^{12}=(\beta^1_2,\beta^2_1)$ running on one machine. Thus when node 1 and node 2 agree that planets 1 and 2 are close, but that planet 3 is further away, an attempt will be made to start a single process $\beta^{12}$ to model the positions of planets 1 and 2. Should that attempt fail (for example a necessary communications link not be established), then the separate processes $\beta^1_2$ and $\beta^2_1$ will continue to run. 

The cooperative process $\beta^{12}$ ends when we leave $\mathcal{M}^{12}\subset \mathcal{M}^{1}\times \mathcal{M}^{2}$. The 
process $\beta^{12}$ makes sure that the pictures $\mathrm{State}^1_{\beta^1_2}$ and $\mathrm{State}^2_{\beta^2_1}$ in nodes 1 and 2 respectively are updated, and then switches control back to the individual nodes in modes $(\beta^1_2,\beta^2_1)\in  \mathcal{M}^{1}\times \mathcal{M}^{2}$.
Given that $(\beta^1_2,\beta^2_1)$ is no longer the preferred mode, either node 1 will then instigate a mode transition, or node 2 will do so. The cooperative mode transfers control to $(\beta^1_2,\beta^2_1)\in  \mathcal{M}^{1}\times \mathcal{M}^{2}$ simply because it was not necessary to duplicate the machinery of the mode transition functions. 

Of course, we should also have $\mathcal{M}^{13}\subset \mathcal{M}^{1}\times \mathcal{M}^{3}$ and $\mathcal{M}^{23}\subset \mathcal{M}^{2}\times \mathcal{M}^{3}$ operating on the same principle.
But what happens if all three planets are close? Following from two node cooperations, we could have three node cooperations 
$\mathcal{M}^{123}\subset \mathcal{M}^{1}\times \mathcal{M}^{2}\times \mathcal{M}^{3}$, consisting of just one mode
$\gamma^{123}=(\gamma^1_{23},\gamma^2_{13},\gamma^3_{12})$.

\section{Concluding remarks and further directions}\label{Concluding_remarks}

We have introduced the idea of a general analogue-digital system with many physical modes and created a data type, the nerve, to manage the relationship between the modes. We have discussed how modes arise and cover -- or fail to cover -- every physical eventuality. Finally, we return to the comparison of our approach with some established formal methods, and comment on the some further refinements to do with classic questions about parallelism, hierarchy and security.

\subsection{Comparison with hybrid systems}\label{HybridSystemsConclusion}

Our fresh look at analogue-digital systems introduces some new ideas for theorising hybrid systems. We believe our approach is distinct from the established computational theory of hybrid systems in a number of ways. 
\newline
\newline
\noindent \textit{AD-interface.} We think about systems that combine the analogue with the digital in very broad terms; we are not focussed on applications involving embedded systems, for example. We model from first principles, i.e., developing properties and an agenda from fresh reflections.
\newline
\newline
\noindent \textit{Oracles.} We assume the digital component is computationally universal and can apply the full power of Church-Turing Thesis. This enables us to use computability theory and complexity theory in developing theory. Indeed, a central notion is that of a \textit{physical oracle} to an algorithm, after Turing. 
\newline
\newline
\noindent \textit{Measurement.} We assume the analogue component generates physical measurements for the digital component to process.  Ideally, measurements are continuous and are generated continuously, but the data processed are approximate and usually sampled discretely. Thus, the analogue component delivers streams of approximate data, which involve errors. We explicitly discuss data that may not match the mathematical models used to design software, or arise without any model available.
\newline
\newline
\noindent \textit{Modes.} The systems are sufficiently complicated to require modularisation into distinct modes of physical behaviour. The distinct algorithms therefore need to change modes. Because of errors of measurement, and other factors, this mode switching is a complex nondeterministic phenomenon. We model the nondeterminism by paths through finite dimensional geometric objects, which we call nerves. A control sequence is a path through the modes of the system. The variations in the path are determined by the nerve. In the hypothetical global state space of the system, as physical conditions vary we consciously modify the control path. The nerve is a finite combinatorial representation of this, the higher dimensional structure of the nerve determines the transitions between control sequences.

\subsection{Networks and cooperation} \label{coop}
The two examples of analogue-digital systems with many modes in Sections~\ref{chicane} and \ref{solar} illustrate cooperation between analogue-digital systems: there are two cars and three bodies. In both cases the component systems are identical, but this is not necessary. In a quarry we could have diggers with modes \{digging, moving\} and trucks with modes \{loading, unloading, moving\}. 

Consider a general network of analogue-digital systems and suppose that some of these systems can cooperate; the systems are nodes in the network. There are penalties for cooperation, such as communication speed limits between nodes of the network, or problems with the increased complexity of the system. Designers need to decide under what circumstances to attempt cooperation, and how to proceed if an attempt fails. There has to be a good reason to cooperate between systems, say to speed and simplify the operation of the combined system.

Consider two analogue digital systems $AD^1$ and $AD^2$, with respective set of modes indexed by $\mathcal{M}^1$ and $\mathcal{M}^2$.  For ease of explanation, suppose they are running on different machines. We suppose that these systems are in general unrelated, so that the set of modes for the combined system $\mathcal{M}$ \textit{contains} the product of the modes for the subsystems $\mathcal{M}^1\times \mathcal{M}^2$. If there is no cooperation between $AD^1$ and $AD^2$, we will simply have $\mathcal{M}^1\times \mathcal{M}^2$, with $\mathcal{M}^1$ being executed in parallel to $\mathcal{M}^2$. 

What would a cooperation between systems $AD^1$ and $AD^2$ look like?  Suppose that the designers have identified modes $\alpha^1\in \mathcal{M}^1$ and $\theta^2\in\mathcal{M}^2$ as a possible cooperation. The default is that $\alpha^1\in \mathcal{M}^1$ and $\theta^2\in\mathcal{M}^2$ will still be executed in parallel. 

However, if we are in $(\alpha^1,\theta^2)\in \mathcal{M}^1\times \mathcal{M}^2$, an attempt will be made to start a cooperative mode. This new mode might consist simply of modified versions of $\alpha^1\in \mathcal{M}^1$ and $\theta^2\in\mathcal{M}^2$ which were optimised to run together, or at the other extreme it might consist of a completely new single process taking the place of both $\alpha^1\in \mathcal{M}^1$ and $\theta^2\in\mathcal{M}^2$ and running on a single machine.

\subsection{A multi-scale hierarchy of modes} \label{multi-scale}
What roles do decomposition and modularity, encapsulation and inheritance, play in implementing multi mode analogue-digital systems? The behaviour of the physical system, and therefore the portfolio of models, has plenty of structure. Consider our autonomous car race in subection \ref{carrace}:

Suppose the team decided that it would improve performance and safety to take account of the track conditions, specifically, if the racing track was dry or wet. As the code for mode $\alpha$ (the car is on the track and some distance from the nearest cars) was already quite complicated, they did not add separate cases to it. Instead they split $\alpha$ into two sub-modes, $\alpha$-dry and $\alpha$-wet. The picture $\mathrm{State}_{\alpha\mathrm{-dry}}=\mathrm{State}_\alpha$ and $\mathrm{State}_{\alpha\mathrm{-wet}}$ was \textsf{state}${}_\alpha$ with an additional state variable for just how wet the track was. 
Now only part of the code for $\alpha$ needed to be rewritten to give the  codes for $\alpha$-dry and $\alpha$-wet. 

Further, they decided that the old mode evaluation functions $P_\alpha$ were working perfectly well, so they did not rewrite them. Only if mode $\alpha$ was chosen did a further choice have to be made, between $\alpha$-dry and $\alpha$-wet. They also needed to decide if mode $\alpha$ was to be an \textit{active} mode, i.e.,\ a mode that could control the system, or a purely \textit{passive} mode whose only purpose was to choose between the sub-modes $\alpha$-dry and $\alpha$-wet and to calculate the evaluation functions. They decided that mode $\alpha$ should be active, as it was a perfectly good compromise solution if for some reason neither $\alpha$-dry nor $\alpha$-wet could be chosen reliably.

The answer to the question of decomposition comes from topology.  A \textit{refinement} of a given open cover for a topological space is an open cover -- i.e., collection of open subsets whose union is the whole space -- and so that any subset is completely contained in some subset of the original \textit{coarser} cover. 

A refinement is like increasing the resolution in a picture: we will resolve modes into submodes, an idea to be taken geometrically literally if we compare the nerve of the cover and the nerve of its refinement. 

The set of modes $\mathcal{M}'$ is a \textit{refinement} of $\mathcal{M}$ if for every $\xi'\in \mathcal{M}'$ there is an $\alpha\in \mathcal{M}$ so that $\xi'$ is contained in $\alpha$. We take this containment to be both physical (the subset of the physical state space corresponding to $\xi'$ is a subset of that corresponding to $\alpha$), and virtual (the software for $\xi'$ is inherited from that for $\alpha$). 

In our racing example above, the refinement $\mathcal{M}'$ of the original modes $\mathcal{M}$ of Section~\ref{carrace} contains $\alpha$-dry and $\alpha$-wet, and we have the containments $\alpha$-dry$\subset$$\alpha$ and $\alpha$-wet$\subset$$\alpha$. 

One complication that arises is that of multiple inheritance (see e.g.,\ \cite{muinscdunibl}). If the $\xi'\in \mathcal{M}'$ above is also contained in $\beta\in \mathcal{M}$, then we also need $\xi'$ to inherit from $\beta$. 

In fact, what we have been describing resembles a generalisation of the manifold data type, the idea of pre-sheaf \cite{BredonSheaf}, where data is localised to subsets of a set. Given a collection of subsets of a set, a \textit{presheaf} assigns a structure $\mathcal{P}(U)$ to every $U$ in the collection, together with a map (called the restriction map) $\phi_{VU}:\mathcal{P}(U)\to \mathcal{P}(V)$ for every $V\subset U$ in the collection.\footnote{There are several generalisations of presheaves, which eventually lead to just a functor between categories.} The basic idea is quite simple, given information about a set, we can restrict it to information about a subset. 

The idea of refinement gives an implementation of the idea of presheaf, in that we have a restriction map from \textsf{state}${}_\alpha$ to every $\mathrm{State}_{\xi'}$
for every $\xi'\in\mathcal{M}'$ with $\xi'\subset \alpha$. Such a restriction map is used in going to finer scale structure in a multi-scale hierarchy. In our case we would also have to be able to move to coarser scales, which is not usually implemented in the structure of a presheaf. However one definition in sheaf theory might be useful: the idea of reconstructing data on a coarser resolution by combining finer resolutions. 
A sheaf is a special case of a presheaf where, given subsets $V_i\subset U$ where the union of the $V_i$ is all of $U$, then we can reconstruct $\mathcal{P}(U)$ from the collection of the  $\mathcal{P}(V_i)$.

\subsection{Security} 
One problem with distributed systems is verifying that other parties are who they say they are. We can incorporate security in the data communications without greatly increasing the complexity of the software.\footnote{There might be unintentional wireless interference between systems. There may be an unauthorised physical connection (e.g.\ an unapproved sensor in an autonomous vehicle). There may be a deliberate cyber attack.} As the only part of the system directly connected to external devices, the interface is the obvious place to add security. As the external devices are physically changed or have their software upgraded, so the security software associated with the interface must be altered.

 \unitlength 0.65 mm
\begin{picture}(130,105)(-15,10)
\linethickness{0.3mm}
\put(60,20){\line(0,1){90}}
\linethickness{0.3mm}
\put(60,20){\line(1,0){70}}
\linethickness{0.3mm}
\put(130,20){\line(0,1){90}}
\linethickness{0.3mm}
\put(60,110){\line(1,0){70}}
\linethickness{0.3mm}
\put(65,70){\line(0,1){10}}
\linethickness{0.3mm}
\put(65,80){\line(1,0){25}}
\linethickness{0.3mm}
\put(90,70){\line(0,1){10}}
\linethickness{0.3mm}
\put(65,70){\line(1,0){25}}
\linethickness{0.3mm}
\put(65,60){\line(1,0){25}}
\linethickness{0.3mm}
\put(90,50){\line(0,1){10}}
\linethickness{0.3mm}
\put(65,50){\line(1,0){25}}
\linethickness{0.3mm}
\put(65,50){\line(0,1){10}}
\linethickness{0.3mm}
\put(65,40){\line(1,0){25}}
\linethickness{0.3mm}
\put(0,70){\line(0,1){10}}
\linethickness{0.3mm}
\put(0,70){\line(1,0){30}}
\linethickness{0.3mm}
\put(30,70){\line(0,1){10}}
\linethickness{0.3mm}
\put(0,80){\line(1,0){30}}
\linethickness{0.3mm}
\put(0,50){\line(0,1){10}}
\linethickness{0.3mm}
\put(0,50){\line(1,0){30}}
\linethickness{0.3mm}
\put(30,50){\line(0,1){10}}
\linethickness{0.3mm}
\put(0,60){\line(1,0){30}}
\linethickness{0.3mm}
\put(0,40){\line(1,0){30}}
\linethickness{0.3mm}
\put(30,30){\line(0,1){10}}
\linethickness{0.3mm}
\put(0,30){\line(1,0){30}}
\linethickness{0.3mm}
\put(0,30){\line(0,1){10}}
\linethickness{0.3mm}
\put(65,30){\line(0,1){10}}
\linethickness{0.3mm}
\put(65,30){\line(1,0){25}}
\linethickness{0.3mm}
\put(90,30){\line(0,1){10}}
\linethickness{0.3mm}
\put(100,80){\line(1,0){25}}
\linethickness{0.3mm}
\put(125,30){\line(0,1){50}}
\linethickness{0.3mm}
\put(100,30){\line(1,0){25}}
\linethickness{0.3mm}
\put(100,30){\line(0,1){50}}
\linethickness{0.3mm}
\put(100,90){\line(0,1){15}}
\linethickness{0.3mm}
\put(100,105){\line(1,0){25}}
\linethickness{0.3mm}
\put(125,90){\line(0,1){15}}
\linethickness{0.3mm}
\put(100,90){\line(1,0){25}}
\put(112.5,98){\makebox(0,0)[cc]{query list}}

\put(113,56){\makebox(0,0)[cc]{\textsf{interface}${}_\alpha$}}

\put(78,75){\makebox(0,0)[cc]{security 1}}

\put(78,55){\makebox(0,0)[cc]{security 2}}

\put(78,35){\makebox(0,0)[cc]{security 3}}

\put(14,75){\makebox(0,0)[cc]{device 1}}

\put(14,55){\makebox(0,0)[cc]{device 2}}

\put(14,35){\makebox(0,0)[cc]{device 3}}

\linethickness{0.3mm}
\put(30,75){\line(1,0){35}}
\linethickness{0.3mm}
\put(30,55){\line(1,0){35}}
\linethickness{0.3mm}
\put(30,35){\line(1,0){35}}
\linethickness{0.3mm}
\put(90,75){\line(1,0){10}}
\linethickness{0.3mm}
\put(90,55){\line(1,0){10}}
\linethickness{0.3mm}
\put(90,35){\line(1,0){10}}
\linethickness{0.3mm}
\put(110,80){\line(0,1){10}}

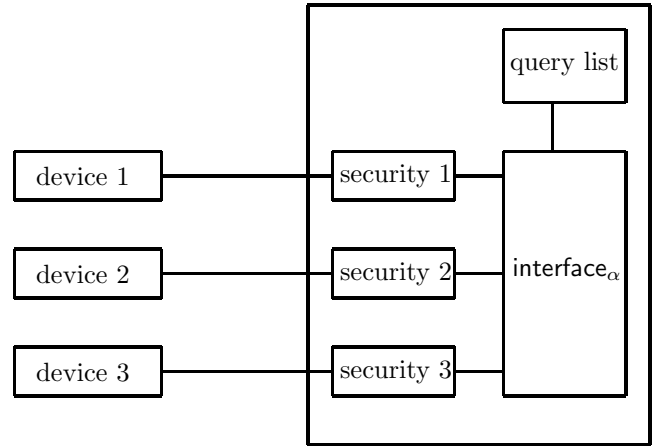
\put(75,0){\makebox(0,20)[cc]{\textbf{Figure 14:} Architecture of the interface with security}}

\end{picture}

%\noindent \textbf{Figure 13:} Architecture of the interface with security

\medskip The idea is to \textit{keep the details of the coding for security and error correction away from the control systems.} However, the control system has to know when its data may be corrupted.  A limited number of flags may be passed from the interface to State${}_\alpha$ describing problems with the data. For example, we might have:

\noindent{\textit{Normal:}}\ Expected behaviour.

\noindent{\textit{Link down:}}\ The device is not answering.

\noindent{\textit{Noise:}}\ More errors in data transmission than normal, probably corrected.

\noindent{\textit{Corrupted:}}\ Errors in data transmission which may not be corrected.

\noindent{\textit{Authentication error:}}\ The device has not authenticated itself properly.

\noindent{\textit{External interference:}}\ The security of the device may be compromised.

\noindent{\textit{Cyber attack:}}\ Possibility that other devices may also be compromised, even if they verify as secure.

\bibliographystyle{plain}

\end{document}